\documentclass[
aps,
prd,
showpacs,
showkeys,
amsmath,
amssymb,
twocolumn,
floatfix,
superscriptaddress]{revtex4-2}

\usepackage{graphicx}
\usepackage{times}
\usepackage{verbatim}
\usepackage{color}
\usepackage{cancel}
\usepackage[normalem]{ulem}
\usepackage{bm}
\usepackage{url}
\usepackage{multirow}
\usepackage{textcomp}
\usepackage{dcolumn}
\usepackage{makecell}
\usepackage{booktabs}
\usepackage{siunitx}
\usepackage{xtab,afterpage}
\DeclareSIUnit{\charge}{\milli\volt\nano\second}
\DeclareSIUnit{\MeV}{\mega\electronvolt}
\DeclareSIUnit{\days}{days}
\usepackage{lineno}
\usepackage{enumerate}
\usepackage{appendix}
\usepackage{tabularx}
\usepackage[colorlinks,linkcolor=blue,anchorcolor=blue,citecolor=blue]{hyperref}
\usepackage{tikz}
\usetikzlibrary{shapes.geometric, arrows}
\usepackage{rotating}
\usepackage[version=3]{mhchem}
\usepackage{footnote}
\usepackage{tikz}
\usetikzlibrary{backgrounds,mindmap,calc,positioning,intersections}
\tikzstyle{startstop} = [rectangle, rounded corners, minimum width = 1cm, minimum height=0.5cm,text centered, draw = black]
\tikzstyle{io} = [trapezium, trapezium left angle=70, trapezium right angle=110, minimum width=1cm, minimum height=0.5cm, text centered, draw=black]
\tikzstyle{process} = [rectangle, minimum width=3cm, minimum height=1cm, text centered, draw=black]
\tikzstyle{decision} = [diamond, aspect = 3, text centered, draw=black]
\tikzstyle{arrow} = [->,>=stealth]

\definecolor{posurl}{cmyk}{.9 .9 0 0}
\usepackage{hyperref}
\hypersetup{
    colorlinks=true
,urlcolor=posurl
,anchorcolor=posurl
,citecolor=posurl
,filecolor=posurl
,linkcolor=posurl
,menucolor=posurl
,linktocpage=true
,pdfa=true
}
\newcommand{\GEANT}{{\textsc{Geant}}}
\newcommand{\Fluka}{{\textsc{Fluka}}}

\usepackage[caption=false]{subfig}

\begin{document}

\title{Study of neutron production for 360 GeV cosmic muons}

\makeatletter
\newcommand{\fmarki}{*}
\newcommand{\fmarkii}{\ensuremath{\dagger}}
\newcommand{\fmarkiii}{\ensuremath{\ddagger}}
\newcommand{\fmarkiv}{\ensuremath{\mathsection}}
\newcommand{\fmarkv}{\ensuremath{\mathparagraph}}
\newcommand{\fmarkvi}{\ensuremath{\|}}
\newcommand{\fmarkvii}{**}
\newcommand{\fmarkviii}{\ensuremath{\dagger\dagger}}
\newcommand{\fmarkix}{\ensuremath{\ddagger\ddagger}}
\def\@fnsymbol#1{{\ifcase#1\or \fmarki\or \fmarkii\or \fmarkiii\or \fmarkiv\or \fmarkv\or \fmarkvi\or \fmarkvii\or \fmarkviii\or \fmarkix \else\@ctrerr\fi}}
\makeatother
\renewcommand{\fmarki}{\ensuremath{\dagger}}

\newcommand{\TUDEP}{\affiliation{Department of Engineering Physics, Tsinghua University, Beijing 100084, China}}
\newcommand{\TUHEP}{\affiliation{Center for High Energy Physics, Tsinghua University, Beijing 100084, China}}
\newcommand{\TUPRI}{\affiliation{Key Laboratory of Particle \& Radiation Imaging (Tsinghua University), Ministry of Education, China}}
\newcommand{\UCASP}{\affiliation{School of Physical Sciences, University of Chinese Academy of Sciences, Beijing 100049, China}}
\newcommand{\SYSUP}{\affiliation{School of Physics, Sun Yat-Sen University, Guangzhou 510275, China}}
\newcommand{\NUSP}{\affiliation{School of Physics, Nanjing University, Nanjing 210093, China}}

\author{Xinshun Zhang}\TUDEP\TUHEP
\author{Jinjing Li}\email[Corresponding author: ]{jinjing-li@mail.tsinghua.edu.cn}\TUDEP\TUHEP
\author{Shaomin Chen}\TUDEP\TUHEP\TUPRI
\author{Wei Dou}\TUDEP\TUHEP
\author{Haoyang Fu}\TUDEP\TUHEP
\author{Ye Liang}\TUDEP\TUHEP
\author{Qian Liu}\UCASP
\author{Wentai Luo}\TUDEP\TUHEP
\author{Ming Qi}\NUSP
\author{Wenhui Shao}\TUDEP\TUHEP
\author{Haozhe Sun}\TUDEP\TUHEP
\author{Jian Tang}\SYSUP
\author{Yuyi Wang}\TUDEP\TUHEP
\author{Zhe Wang}\TUDEP\TUHEP\TUPRI
\author{Changxu Wei}\TUDEP\TUHEP
\author{Jun Weng}\TUDEP\TUHEP
\author{Yiyang Wu}\TUDEP\TUHEP
\author{Benda Xu}\TUDEP\TUHEP\TUPRI
\author{Chuang Xu}\TUDEP\TUHEP
\author{Tong Xu}\TUDEP\TUHEP
\author{Yuzi Yang}\TUDEP\TUHEP
\author{Aiqiang Zhang}\TUDEP\TUHEP
\author{Bin Zhang}\TUDEP\TUHEP

\collaboration{JNE Collaboration}\noaffiliation
\date{\today}

\begin{abstract}
    The China Jinping underground Laboratory (CJPL) is an excellent location for studying solar, terrestrial, and supernova neutrinos due to its 2400-meter vertical rock overburden. Its unparalleled depth gives an opportunity to investigate the cosmic-ray muons with exceptionally high average energy at $\sim360$~GeV. This paper details a study of muon-related backgrounds based on 1178 days of data collected by the 1-ton prototype neutrino detector used for the Jinping Neutrino Experiment (JNE) since 2017. The apparent effects for the leakage of muons' secondary particles due to detector's finite size on the measured neutron yield are first discussed in detail. The analysis of 493 cosmic-ray muon candidates and $13.6\pm5.7$ cosmogenic neutron candidates, along with a thorough evaluation of detection efficiency and uncertainties, gives a muon flux of $(3.56\pm0.16_{\mathrm{stat.}}\pm0.10_{\mathrm{syst.}})\times10^{-10}~\mathrm{cm}^{-2}\mathrm{s^{-1}}$ and a cosmogenic neutron yield of $(3.37\pm 1.41_{\mathrm{stat.}}\pm 0.31_{\mathrm{syst.}}) \times 10^{-4}~\mathrm{\mu}^{-1} \mathrm{g}^{-1} \mathrm{cm}^{2}$ in LAB-based liquid scintillator.
\end{abstract}

\maketitle

\section{Introduction}

Situated in the Jinping Mountains of Sichuan, China, with a vertical rock overburden of about 2400 meters, the China Jinping Underground Laboratory (CJPL) is one of the world's deepest underground laboratories~\cite{Cheng:2017usi}. This unparalleled depth provides exceptional shielding from cosmic rays, which is crucial for detecting solar neutrinos and various rare signals~\cite{Zeng:2020cyw, su2012}.

The Jinping Neutrino Experiment (JNE) aims to study MeV-scale neutrinos, such as solar, terrestrial, and supernova neutrinos, whose measurements are highly sensitive to muons and the cosmogenic radioactive backgrounds~\cite{Jinping:2016iiq}. For example, in organic scintillators, $^{11}\mathrm{C}$ can be produced by cosmic muons through spallation processes on $^{12}\mathrm{C}$. A detailed discussion of the related processes can be found in Ref.~\cite{Galbiati:2004wx}. The $\beta^+$-decays of $^{11}\mathrm{C}$ are largely indistinguishable from neutrino-induced electron recoils, presenting a challenge for the observation of Carbon-Oxygen-Nitrogen and $pep$ solar neutrinos~\cite{Borexino:2013zhu,Borexino:2021pyz, BOREXINO:2020hox}. The cosmic-ray spallation products, such as $^{9}\mathrm{Li}$, $^{11}\mathrm{Be}$ and $^{16}\mathrm{N}$, also act as an important background in searches for the upturn of the solar $^{8}\mathrm{B}$ neutrino spectrum~\cite{Super-Kamiokande:2023jbt, Borexino:2017uhp} and supernova relic neutrinos~\cite{Super-Kamiokande:2015xra}. Among those products, cosmogenic neutrons, created in nuclear interactions triggered by cosmic muons, are significant backgrounds for many rare-event searches, even in deep underground detector~\cite{SNO:2019pzy}. Furthermore, the muon flux can differ significantly between laboratories located under mountains and those below mine shafts with the same vertical rock overburden~\cite{Mei:2005gm}. Therefore, it is important to investigate the production of cosmogenic neutrons and cosmic muon flux at CJPL, as the depth results in significantly elevated average muon energy.

A 1-ton scintillator detector, built as a prototype for the JNE, had been operational since 2017 at the first construction phase of CJPL (CJPL-I). The prototype was decommissioned and disassembled on September 3, 2023, for further upgrade. This detector was designed to evaluate the performance of key components and technologies, and measure the underground background levels in situ~\cite{Wu:2022oxo}.

Previous studies provide measurements of cosmic muon flux and cosmogenic neutron yield in linear alkylbenzene (LAB)-based liquid scintillator at CJPL-I using part of the data collected by the detector~\cite{JNE:2020bwn, JNE:2021cyb}. This study utilizes the complete data set to give new measurements of cosmic muon flux and cosmogenic neutron yield with improved evaluations of efficiencies, corrections and uncertainties, providing a comprehensive understanding of cosmic-ray muon and cosmogenic neutrons with the mountain's topography around the laboratory. Furthermore, this study first gives a detailed study of the effect induced by detector's finite-size in the measurement of cosmogenic production yield.

Following the detailed description of the design and operational aspects of the 1-ton prototype provided in Section~\ref{sec:Experiment}, Section~\ref{sec:Simulation} presents the simulations undertaken to model the passage of muons through the Jinping mountain and their interactions within the detector. In Section~\ref{sec:DataAnalysis}, we outline the preliminary data analysis procedures, including data quality assessments and the methodologies employed for event reconstruction. Section~\ref{sec:Muons} delves into the muon candidate selection and reconstruction. The analysis details and measurement result of cosmogenic neutron yield are discussed in Section~\ref{sec:Neutrons}. Finally, Section~\ref{sec:Summary} provides a summary of the findings and discusses prospects.

\section{The 1-ton Prototype}\label{sec:Experiment}
\begin{figure}[!htbp]
    \centering
    \includegraphics[width=0.95\columnwidth]{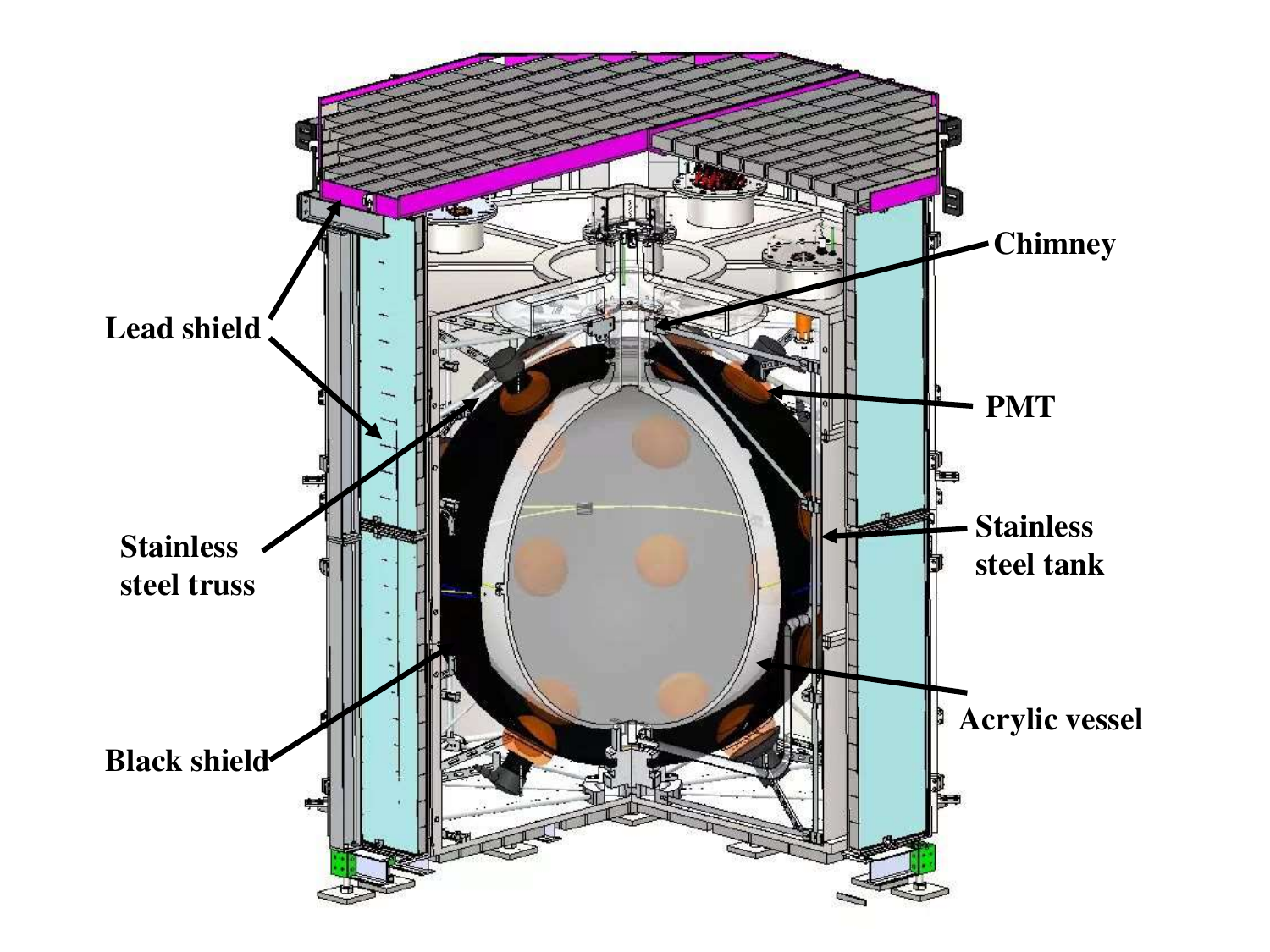}
    \caption{The schematic figure of the 1-ton prototype for JNE.}
    \label{fig:detector}
\end{figure}

Fig.~\ref{fig:detector} depicts the structural design of the 1-ton prototype~\cite{Wang:2017ynm}. The central target volume was encompassed by a 645-mm-radius spherical acrylic vessel with a thickness of 20~mm. This vessel contained the slow liquid scintillator (LS), which was composed of LAB as a solvent, doped with 0.07~g/L of the fluor 2,5-diphenyloxazole (PPO) and 13~mg/L of the wavelength shifter 1,4-bis(2-methylstyryl)benzene (bis-MSB). The LS can emit scintillation light with an extended duration of about 30~ns, which helps in separating Cherenkov light from scintillation light event by event~\cite{Guo:2017nnr, Luo:2022xrd}. This feature is crucial for the directional measurement in LS detector~\cite{BOREXINO:2021xzc, BOREXINO:2021efb, SNO:2023cnz}.

The target volume was encompassed by pure water held within a stainless steel tank that is 4~mm thick, and had a diameter of 2000~mm, and a height of 2090~mm. Within the water, there were thirty 8-inch Hamamatsu R5912 photomultiplier tubes (PMTs) oriented inward to detect the photons emitted from the LS. The water served as a passive shield against external radiation. The tank was further enclosed by a 5-cm-thick lead wall, offering additional shielding to suppress ambient radioactive background.

After two-year data collection period, we observed the $\beta$-$\alpha$ cascade decay of $^{214}\mathrm{Bi}$-$^{214}\mathrm{Po}$ near the chimney at the top indicating the external radon gas leakage into the target volume. To shield against this background, we installed a nitrogen bubbling system between July 11 and 14, 2019, to maintain positive pressure inside the detector, thus preventing air infiltration into the acrylic vessel. This system offered us an additional benefit - an increase of light yield in the LS, as the degassing process with nitrogen effectively removes dissolved oxygen from the LS, thereby reducing the quenching effect.

The front-end electronic system consisted of four CAEN V1751 FlashADC boards and a CAEN V1495 logical trigger module. Each FlashADC board had eight channels, 10-bit ADC precision for a 1~V dynamic range, and a 1~GHz sampling rate. All PMT signals were directly fed into the FlashADC boards for digitization. If more than $N_{\mathrm{PMT}}$ PMTs are triggered with a threshold $H_{\mathrm{th}}$, the data acquisition (DAQ) system record the pulse shapes of all the activated PMTs with a $T_{\mathrm{w}}$-ns waveform sampling length. The specific values are listed in Table~\ref{tab:trigger-conditions}. Further details about the detector system can be found in Ref.~\cite{Wu:2022oxo}.
\begin{table*}[!htbp]
    \tabcolsep=0.45cm
    \centering
    \caption{The details of each DAQ phase for the 1-ton prototype. $T_\mathrm{DAQ}$: effective DAQ time. $N_{\mathrm{PMT}}$: the least number of the fired PMTs for trigger. $T_{\mathrm{w}}$: sampling length of the waveform in a single channel. $H_{\mathrm{th}}$: trigger threshold of a single channel in unit of the fraction of the single photoelectron signal amplitude. Nitrogen bubbling: presence of nitrogen bubbling system. $R$: average event rate.}\label{tab:trigger-conditions}
    \begin{tabularx}{\textwidth}{cccccccc}
        \hline
        \hline
        Start Date & End Date   & ${T_{\mathrm{DAQ}}}$~[day] & $N_{\mathrm{PMT}}$ & $T_{\mathrm{w}}$~[ns] & $H_{\mathrm{th}}$ & Nitrogen Bubbling & $R$~[Hz] \\
        \hline
        2017-07-31 & 2018-10-14 & 392.0                      & 25                 & 1029                  & 43\%              & No                & 28.8     \\
        2018-10-15 & 2019-06-29 & 238.5                      & 25                 & 1029                  & 22\%              & No                & 33.3     \\
        2019-06-30 & 2019-07-14 & 12.6                       & 10                 & 600                   & 22\%              & No                & 135.9    \\
        2019-07-15 & 2023-09-02 & 534.9                      & 10                 & 600                   & 22\%              & Yes               & 158.9    \\
        \hline
        \hline
    \end{tabularx}
\end{table*}

\section{Simulation}\label{sec:Simulation}
To evaluate the detection efficiencies of cosmic-ray muons and cosmogenic neutrons, we need to understand their responses in the detector. We have developed a \GEANT4-based Monte Carlo (MC) simulation framework to analyze the muon interaction processes within the surrounding mountain rock and the detector materials~\cite{GEANT4:2002zbu,allison2006geant4}. The simulations are divided into two main parts: mountain-related simulation and detector-related simulation. The first one gives the cosmic-ray muons' energy and angular distributions as they arrive at the underground laboratory, with the surrounding mountain terrain incorporated. The second one uses a muon generator informed by the mountain simulation results, producing processed PMT waveforms from the electronics for each event, with features close to the real data.

\subsection{Mountain Related Simulation}
The mountain simulation uses two primary muon generators: the modified Gaisser's formula, which works well at low energies~\cite{guan2015parametrizationcosmicraymuonflux}, and MCEq, which numerically calculates the propagation and interactions of cosmic rays in the atmosphere~\cite{Fedynitch:2015zma,Fedynitch:2018cbl}. In this study, we use the results from the MCEq simulation as the nominal ones and take the differences between the two generators as systematic uncertainties.

\begin{figure*}[!htbp]
    \centering
    \subfloat[]{\includegraphics[width=0.48\linewidth]{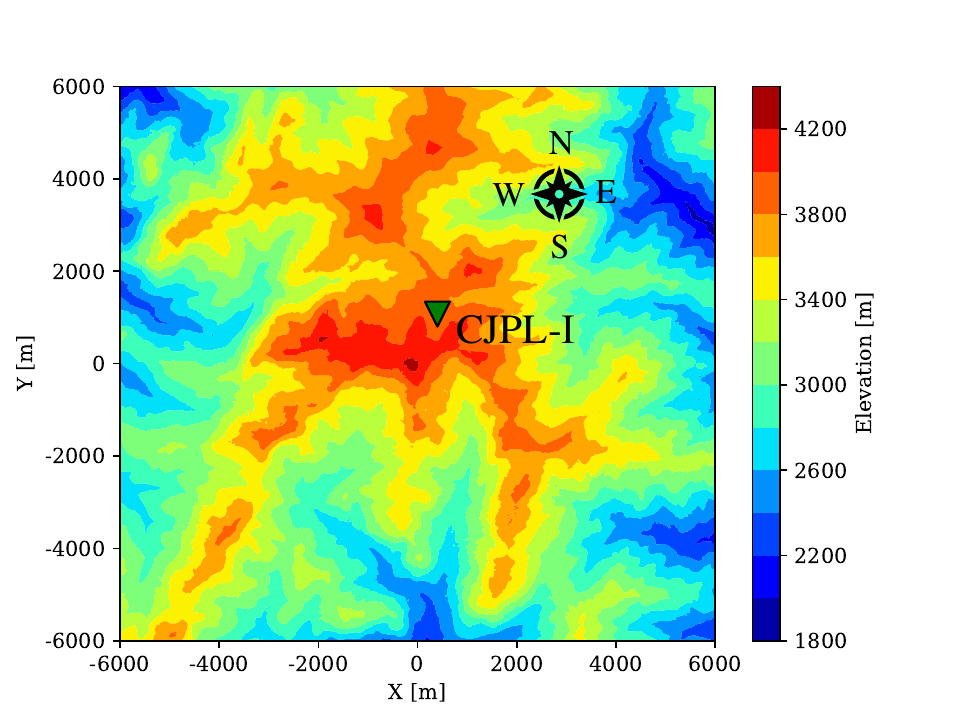}}
    \subfloat[]{\includegraphics[width=0.48\linewidth]{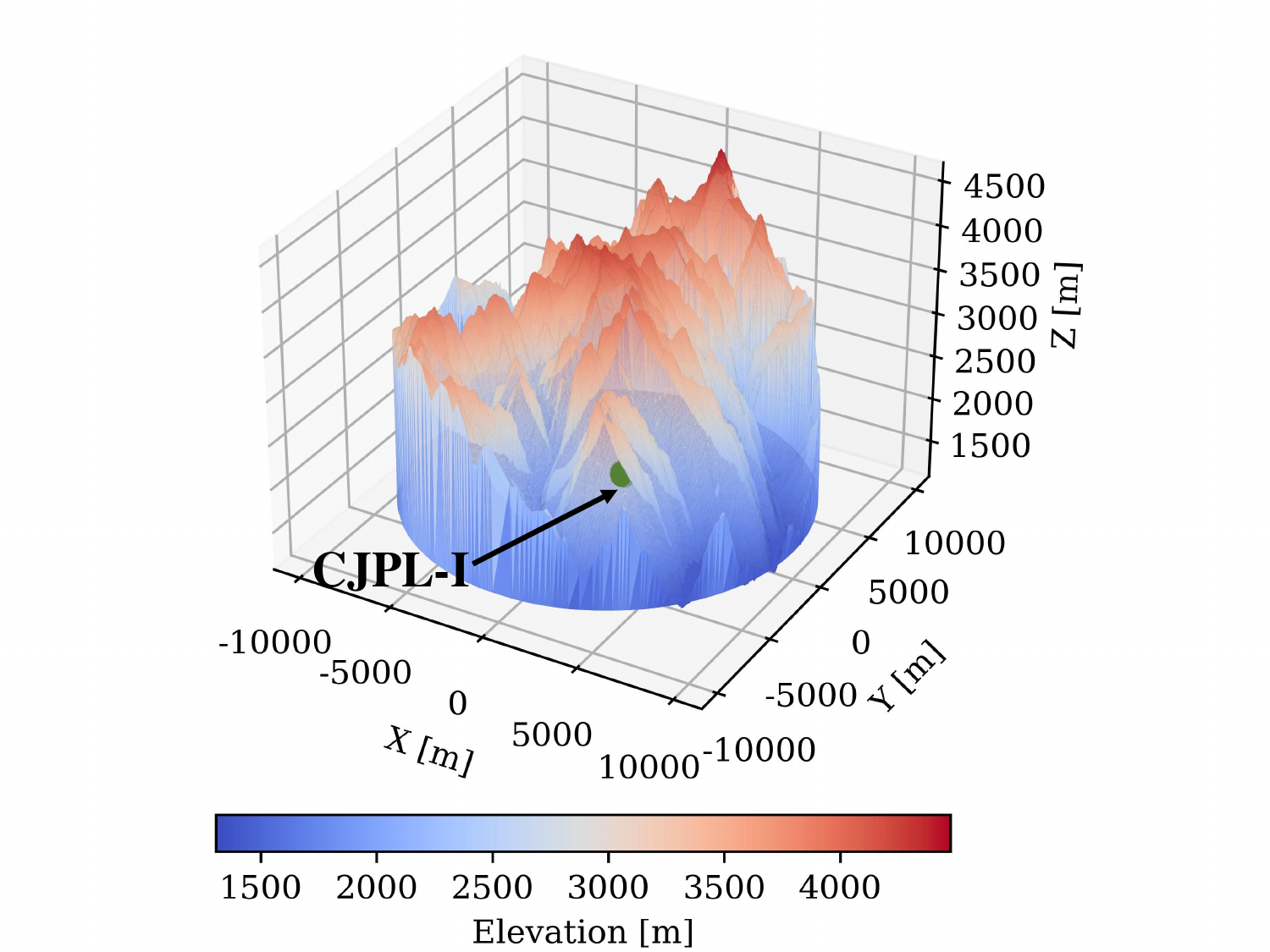}}
    \caption{(a): Contour map of mountain terrain near CJPL, given by the NASA SRTM3 dataset~\cite{terrain2007}. (b): The mountain geometry obtained from the Delaunay Triangulation of mountain terrain in the mountain simulation. The positive direction on the x-axis corresponds to true East, and the positive direction on the y-axis corresponds to true North. The z-axis is the absolute elevation. The green solid point represents the position of CJPL-I.}
    \label{fig:contour}
\end{figure*}

\begin{figure}[!htbp]
    \centering
    \subfloat[]{\includegraphics[width=0.95\columnwidth]{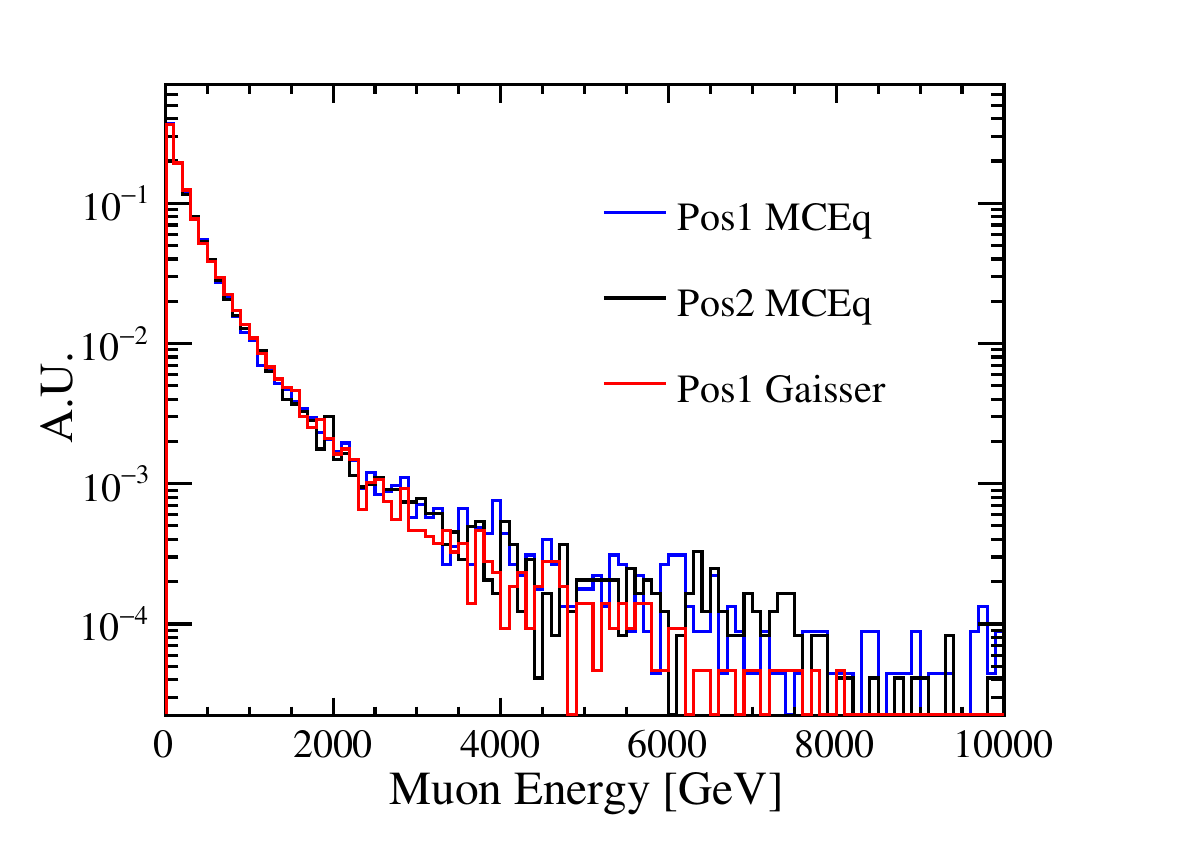}}\hspace*{\fill}\\
    \subfloat[]{\includegraphics[width=0.95\columnwidth]{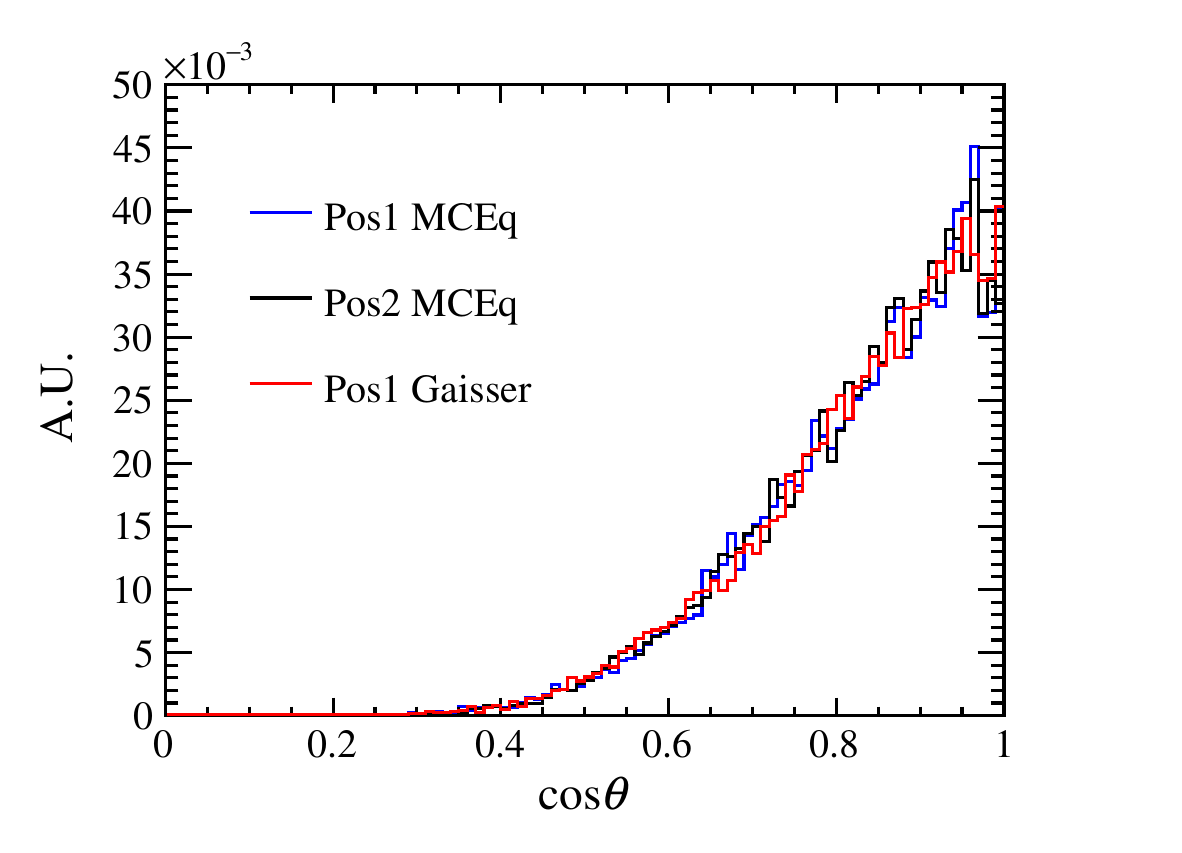}}\hspace*{\fill}\\
    \subfloat[]{\includegraphics[width=0.95\columnwidth]{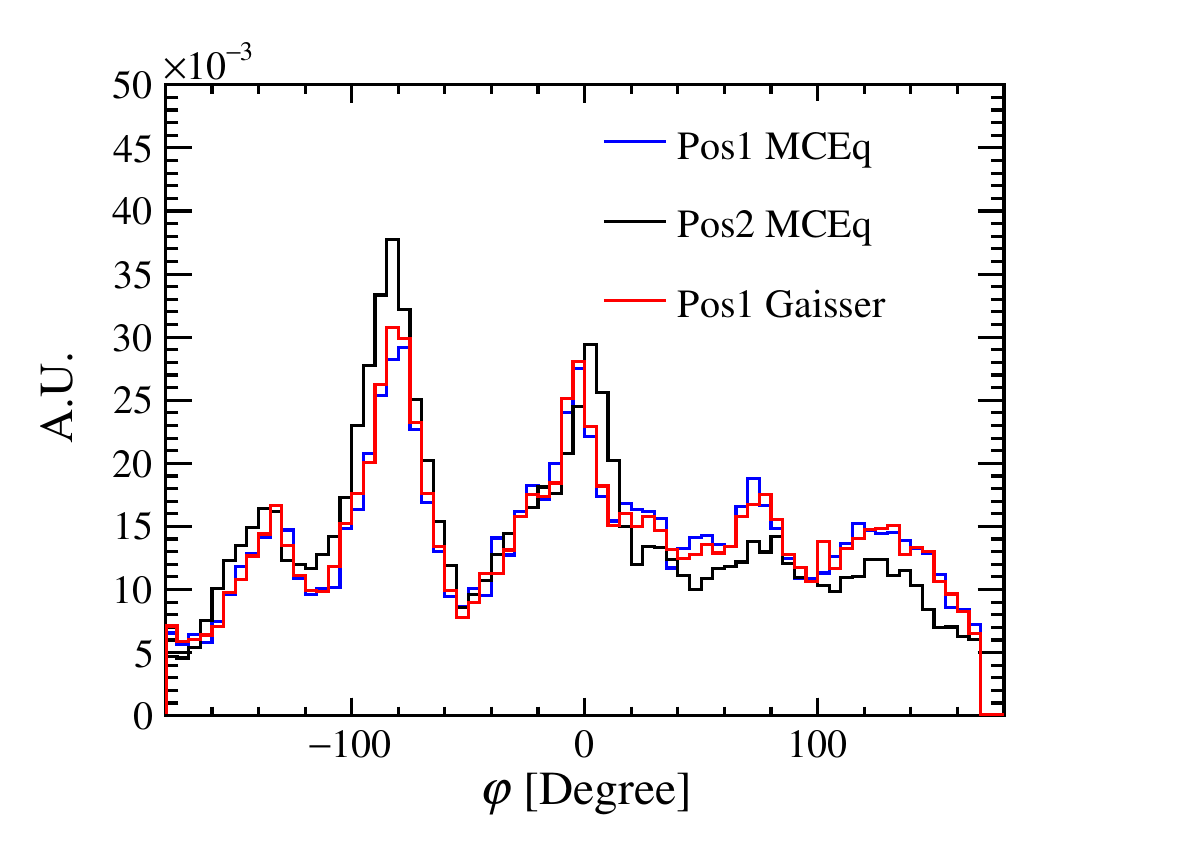}}

    \caption{The energy and angular distributions of muons arriving at the underground laboratory derived from the simulation. $\theta$ represents the zenith angle, defined as the angular distance from the vertical, while $\varphi$ denotes the azimuth angle, defined as the angular distance from true East in the horizontal plane. Pos1 corresponds to the location of CJPL-I, and Pos2 is situated 100 meters south of CJPL-I. Two primary muon generators, MCEq and the modified Gaisser's formula are used.}
    \label{fig:generator}
\end{figure}

The terrain data of the Jinping mountain region is sourced from the NASA SRTM3 dataset~\cite{terrain2007}. The mountain terrain is accurately integrated into the \GEANT4 geometry using the Delaunay Triangulation method. This method can group discrete elevation points into triangles to model the mountain surface. Two figures in Fig.~\ref{fig:contour} show the contour map and the 3D representation of the mountain structure near CJPL-I. In simulation, the mountain is assumed to consist of a uniform density of rock set to 2.8~$\mathrm{g/cm}^3$~\cite{zheng2024three}. The rock composition is $46.1\%$ oxygen, $28.2\%$ silicon, $8.2\%$ aluminum, and $5.6\%$ iron, aligning with the typical abundance found in Earth's crust~\cite{earth2016}. Fig.~\ref{fig:generator} shows the simulated energy and angular distributions of muons arriving at CJPL-I.

\subsection{Detector Related Simulation}
The detector-related simulation, including the complete structure of the 1-ton prototype, utilizes the mountain-related simulation results as input. The waveform saturation effect due to the significant particle energy deposits is also simulated. The same analysis procedures used in data, are applied to the simulation data to estimate efficiencies.

\begin{figure}[!htbp]
    \centering
    \includegraphics[width=0.95\columnwidth]{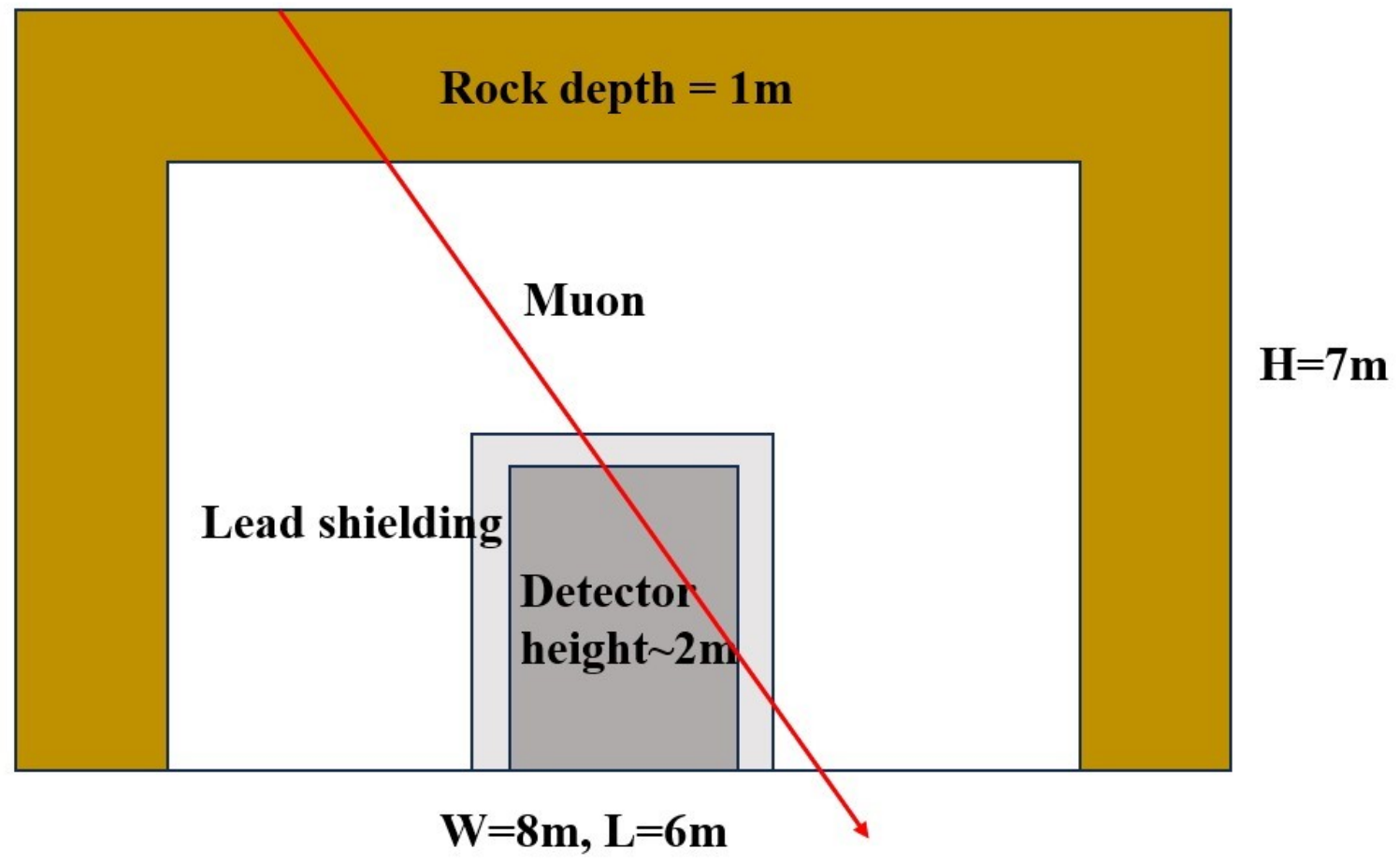}
    \caption{The geometry setup in the detector related simulation.}
    \label{fig:detector-simulation-structure}
\end{figure}

Fig.~\ref{fig:detector-simulation-structure} shows a schematic diagram of the geometry setup for the detector simulation. Outside the detector, a 1-meter-thick layer of rock is added to account for the muon shower, which can affect the measurement of cosmic-ray muon flux and cosmogenic neutron yield. In the detector simulation, muons are generated in the rock as shown in Fig.~\ref{fig:detector-simulation-structure}, according to the input distribution. The corresponding detector responses are recorded for the evaluation of efficiencies and uncertainties in the measurement.

All tracks of secondary particles involved in neutron production are recorded to correct the number of measured cosmogenic neutrons in the LS. Neutrons are generated by muons dominantly in hadron and electromagnetic showers~\cite{Li:2015kpa}. Three physics lists in \GEANT4~(version 10.06.p03) are employed to address this process: QGSP-BIC-HP, QGSP-BERT-HP, and FTFP-BERT-HP. The Quark-Gluon String model (QGS) and the FRITIOF String model (FTF) are used for hadronic interactions at high energies ($>10$~GeV), while the Binary Cascade model (BIC) and the Bertini Cascade model (BERT) are applied for hadronic interactions at lower energies (between 70~MeV and 9.9~GeV). The physical processes below 70~MeV and nuclear de-excitation are simulated using \GEANT4's precompound model. A high-precision neutron model (HP) is used for neutron elastic and inelastic interactions below 20~MeV. The uncertainties using different physics lists are evaluated by comparing these models and will be discussed in Sec.~\ref{sec:Neutrons}.

The simulation results will be compared to the measurements in the following sections.

\section{Data Analysis}\label{sec:DataAnalysis}
\subsection{Data Sets}
This study uses the complete dataset collected by the detector from July 31, 2017, to September 2, 2023. Table~\ref{tab:trigger-conditions} summarizes the detailed run operations and conditions.

\subsection{Data Quality}
Data quality is ensured through a two-step checking process, with the minimal unit being a $\sim$200~MB file typically spanning about 5 minutes under the final trigger condition. However, the exact duration depends on the actual event rate as shown in Table~\ref{tab:trigger-conditions}.

The first step is to manually select physics runs, excluding problematic runs identified from issues recorded during the shift, such as pedestal calibration errors, detector maintenance, hardware issues, or external disturbances.

The second step is to identify bad channels based on baseline fluctuations and channel occupancy. Baseline fluctuation is quantified by the standard deviation of the baseline, initially calculated from the first 150~ns of the waveform before the main pulse. It is changed to 30~ns after changing $T_\mathrm{w}$ as shown in Table~\ref{tab:trigger-conditions}. Channels with fluctuations exceeding 2.5~mV are marked as bad channels. Given that the mean baseline fluctuation for a normal channel is less than 1~mV, the 2.5-mV threshold effectively identifies channels impacted by electronic noise or detector malfunctions. Channel occupancy is defined as the ratio of the number of triggers for a channel to the total number of trigger events in a unit file. Due to the detector's spherical symmetry and the uniform distribution of the PMTs, channel occupancies are expected to be uniform. Any channel with an occupancy that deviates from the mean by more than 5~$\sigma$ is tagged as a bad channel. Events in unit files with no more than four bad channels are included in the analysis but require reconstruction with bad channel corrections. Unit files with more than four bad channels are excluded from subsequent analysis.

These checks exclude about 4.9\% of the data, resulting in an effective DAQ time of $\sim1178.0$ days for subsequent analysis.

\subsection{PMT Calibrations}
The gain and time calibrations of PMT rely on the dark noise of the PMTs and the natural decay products of radioisotopes within the detector, respectively, as detailed in Ref.~\cite{Wu:2022oxo}.

A real-time gain calibration is performed using the ``RollingGain'' method. This method calculates the charge distribution of dark noise within a given period. It is fitted using a probability distribution function that accounts for the signal generated by thermionic emission electrons from the photocathode and potential background noise~\cite{bellamy1994absolute}. The gain factor for each channel during this period is derived from the fitting result and is used to convert the measured charge into the number of photoelectrons (PE). By accumulating PEs from the good channels, the deposited energy is reconstructed in terms of the PE numbers.

The time delay between the PMT hits and the readout varies channel by channel. Therefore, time calibration is crucial when determining the sequence of PMT hit times for event reconstruction. The sample for the calibration is selected by requiring a relatively large PE number and vertex close to the detector center. The time calibration finally provides the relative offsets among different channels. The typical value is within $\pm$5~ns.

\subsection{Energy Scale}
We utilize the 2.61-MeV $\gamma$ peak from $^{208}\mathrm{Tl}$ decay originates from the natural contamination inside detector to calibrate the energy scale. The fitted peak position gives the corresponding PE number, which can be used to determine the nominal energy scale calibration factor. This factor is determined to be 61.07~PE/MeV before July 14, 2019, and then increases to 88.02~PE/MeV due to the nitrogen bubbling between July 15, 2019, and July 22, 2019. Finally, it becomes stable at 99.89~PE/MeV. To cross-validate these values, we further fit the spectrum of $\alpha$ particles from $^{214}\mathrm{Po}$ decay, and the 1.46-MeV $\gamma$ rays from $^{40}\mathrm{K}$ decay to evaluate the calibration factor independently. The differences between the nominal energy scale factors and those independently derived from $^{40}\mathrm{K}$ remain within 5\% across the entire dataset, and this variation is treated as a systematic uncertainty. The resulting visible energy of the $\alpha$ from $^{214}\mathrm{Po}$ is determined to be approximately 0.85~MeV, which is comparable to that presented in the Daya Bay experiment~\cite{an2017measurement}. Further details on the energy scale calibration are available in Ref.~\cite{Wu:2022oxo}.

\section{Cosmic Ray Muon}\label{sec:Muons}
\subsection{Event Selection}
In this analysis, all the muon candidates are required to have a visible energy exceeding 60~MeV. At this energy level, the primary backgrounds include spontaneous light emission from PMTs (flashers) and electronic noise, which can be excluded by further analysis of the charge pattern and waveform structure.

When a flasher occurs, neighboring PMTs are also likely to be illuminated. Its charge distribution is less uniform across channels than muon events, with the discharge channel collecting a significantly higher charge than others. We define a parameter $r_{\mathrm{max}} \equiv P_{\mathrm{max}} / \sum_{i=1}^{30}P_{i}$ to identify this background, where $P_{\mathrm{max}}$ is the maximal PE number, and $P_{i}$ is the PE number of channel $i$, calibrated by the gain factor. Events with large $r_{\mathrm{max}}$ tend to be flashers.

Electronic noises caused by accidental external disturbances to the detector's electronic system can lead to extreme baseline fluctuations across all channels. These fluctuations can generate triggers, mimicking high-energy signals. The mean peak number $n_{\mathrm{peak}}$ of the waveforms from the good channels in an event is used to identify electronic noises. For normal events, the peak number is related to the arrival photons registered by the PMTs and is approximately proportional to the deposited energy. However, the peak number for electronic noise is significantly higher than a normal event can have.

\begin{figure}[!htbp]
    \centering
    \includegraphics[width=0.95\columnwidth]{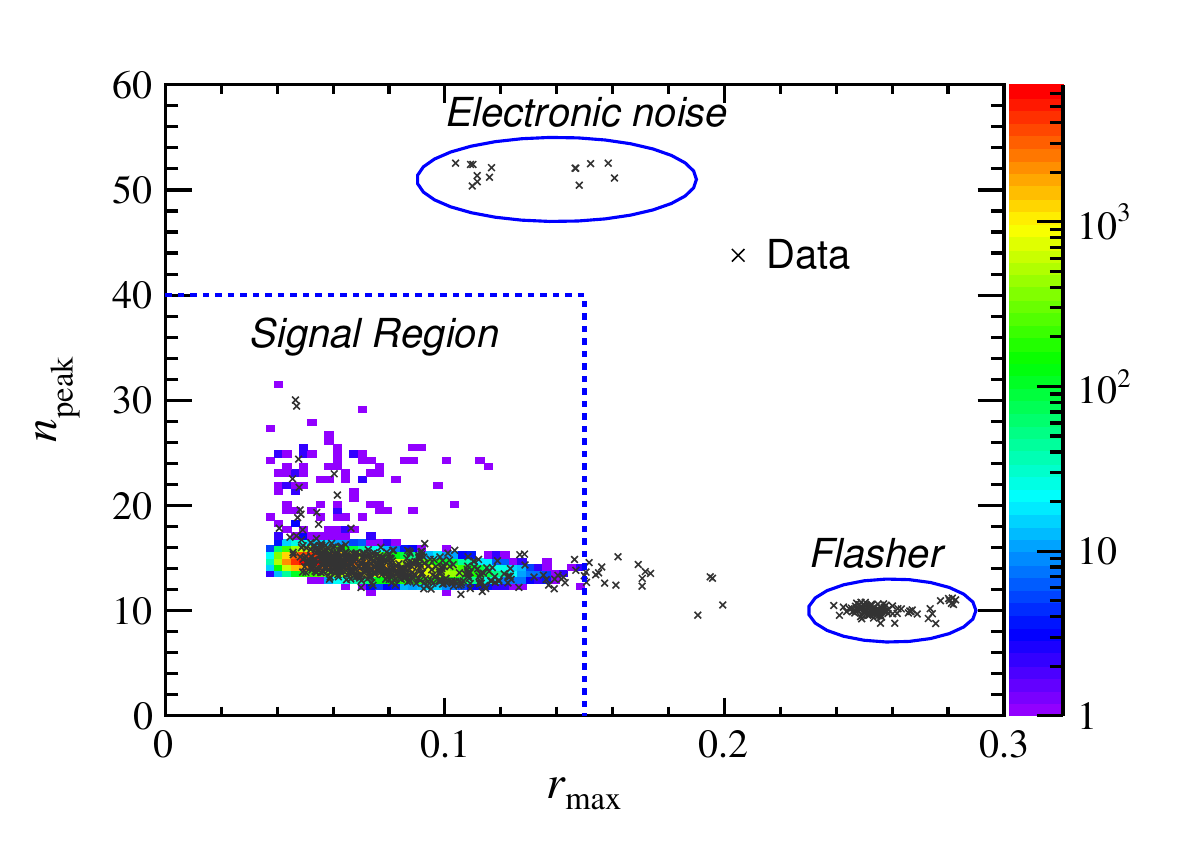}
    \caption{$r_{\mathrm{max}}$ versus $n_{\mathrm{peak}}$ in the data (black cross) and simulation (color block). The color bar represents the number of counts in the simulation. Typical muon candidates are in the region for $r_{\mathrm{max}} <0.15$ and $n_{\mathrm{peak}} < 40$, while flasher and electronic noise events have larger $r_{\mathrm{max}}$ and $n_{\mathrm{peak}}$, which are distributed in the blue circled region.}
    \label{fig:muon-selection}
\end{figure}

The distribution of $r_{\mathrm{max}}$ versus $n_{\mathrm{peak}}$ is shown in Fig.~\ref{fig:muon-selection}. By comparing muon events in the detector simulation with those in the data, we require $r_{\mathrm{max}} < 0.15$ to suppress flashers and $n_{\mathrm{peak}} < 40$ to reject electronic noise. These selection criteria are validated using the MC sample and taken into consideration when evaluating the efficiencies.

\begin{figure}[!htbp]
    \centering
    \includegraphics[width=0.95\columnwidth]{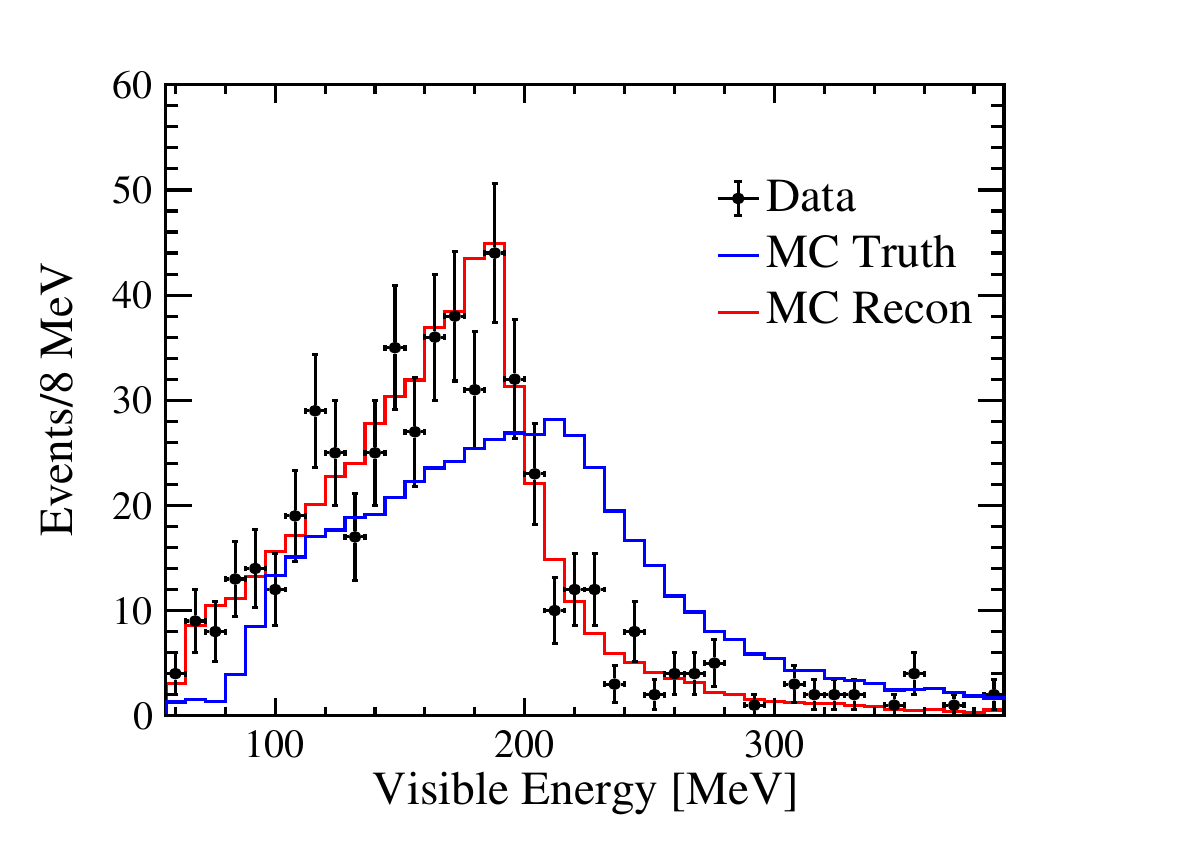}
    \caption{The visible energy distribution of selected muon events in data (block dots) and simulation. The blue line represents the true deposited energy of muons in MC sample while the red line shows the reconstructed energy of muons in MC sample. The differences between the MC truth and MC Recon is induced mainly by the waveform saturation.}
    \label{fig:muon-energy}
\end{figure}

This analysis selects 493 muon candidates, corresponding to an observed event rate of 0.42~events per day. Fig.~\ref{fig:muon-energy} shows distribution of measured visible energy for muons in data and MC sample. Due to the waveform saturation and the energy non-linearity effect, the reconstructed energy of muons in the detector is less than the deposited one, which is especially significant for high-energy events. The average energy of muons detected in this study is investigated using \GEANT4 simulations, yielding approximately 340~GeV before and 360~GeV after applying the selection criteria. The latter value, which differs from the former one reported in the previous study~\cite{JNE:2021cyb}, is used in this work.
\subsection{Direction Reconstruction}
\begin{figure}[!htbp]
    \centering
    \includegraphics[width=0.95\columnwidth]{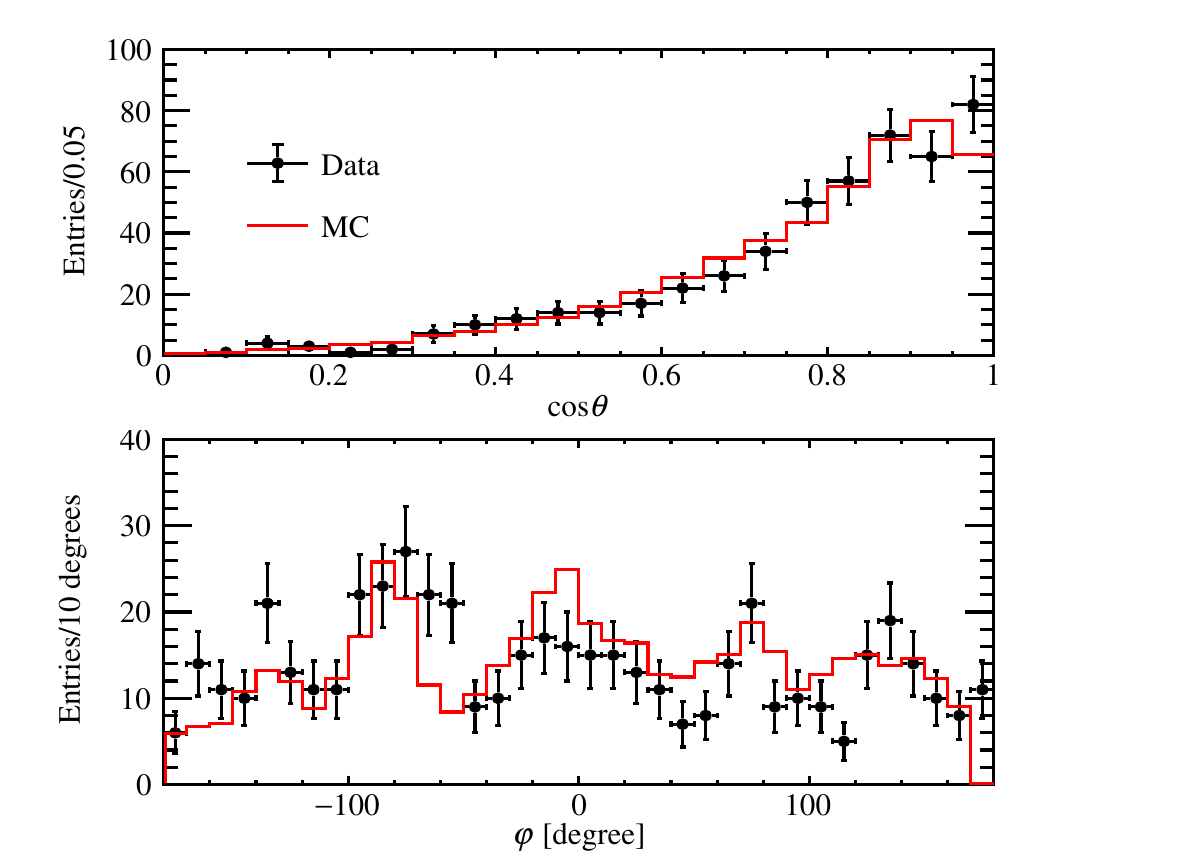}
    \caption{The reconstructed muons' angular distributions in the data (black dots) and simulation (red line). $\theta$ and $\varphi$ correspond to the zenith and azimuth angles as defined before, respectively.}
    \label{fig:angle}
\end{figure}

This study uses a template-based method to reconstruct the muon direction~\cite{JNE:2020bwn}. The muon templates are generated from the detector simulation, each tagged with the direction, and the entry point on the acrylic vessel. The direction is first sampled uniformly across all solid angles. Then the entry point on the vessel surface are sampled uniformly from a hemisphere facing the muon direction. For the generated template and selected event in data, the PMT arrival time vectors are both constructed and zero-centered by subtracting the mean value. The Euclidean distance between the vectors in templates and data is calculated to quantify the similarity. For the event under reconstruction, the $l$-closest templates in Euclidean distance are selected. Finally, the corresponding $l$ directions of the templates are weighted by the reciprocal of the distance to estimate the muon direction. The reconstructed distributions of the azimuth and zenith angles for both data and simulation are in good consistency, as shown in Fig.~\ref{fig:angle}, reflecting the mountain structure above CJPL-I. The small difference between data and simulation may result from the cavities inside the mountain like the tunnels~\cite{zhang2022mou}.

\subsection{Detection Efficiency}
The overall efficiency, $\varepsilon$, is the ratio of muons passing the selection criteria described above to the total number of muons arriving at the experiment hall. It is investigated by the simulations described in Sec.~\ref{sec:Simulation}. Besides the muons passing directly through the detector, those interacting with the surrounding rock and lead shielding can produce showers with secondary high-energy particles, affecting the efficiency. Therefore, the entire experiment hall is considered in the simulation to account for this effect. We decompose the efficiency $\varepsilon$ into three components: the geometry efficiency $\varepsilon_\mathrm{g}$, the detection efficiency $\varepsilon_\mathrm{d}$, and the shower efficiency $\varepsilon_\mathrm{s}$, as
\begin{equation}
    \varepsilon = \varepsilon_\mathrm{g}\times\varepsilon_\mathrm{d}+\varepsilon_\mathrm{s},
\end{equation}
\begin{equation}
    \varepsilon_\mathrm{g} = \frac{N_\mathrm{p}}{N_\mathrm{total}}, \varepsilon_\mathrm{d} = \frac{N_\mathrm{d}}{N_\mathrm{p}}, \varepsilon_\mathrm{s} = \frac{N_\mathrm{s}}{N_\mathrm{total}},
\end{equation}
where $N_\mathrm{total}$ is the total number of muons arriving at the experimental hall, $N_\mathrm{p}$ is the number of muons passing through the sensitive volume of the detector, $N_\mathrm{d}$ is the number of detected muons passing the selection criteria, $N_\mathrm{s}$ is the number of detected muon showers passing the selection criteria.

According to the detector simulation mentioned above, the efficiencies are calculated to be $\varepsilon = 1.7\%, \varepsilon_g = 2.1\%,\varepsilon_d = 82.2\%$ and $\varepsilon_s = 0.3\text{\textperthousand}$, respectively.

\subsection{Result and Uncertainties of Muon Flux Measurement}
The cosmic muon flux at CJPL-I can be calculated as
\begin{equation}
    \phi_\mu=\frac{N_{\mathrm{total}}}{T\times S}=\frac{N_\mu}{\varepsilon \times S \times T},
\end{equation}
where $N_\mu$ is the number of selected candidates in data, $T$ is the effective DAQ time, $S$ is the equivalent projection area of experiment hall and is calculated to be 78.7~$\mathrm{m}^2$ at CJPL-I using the angular distribution of underground muons as detailed in Ref.~\cite{JNE:2020bwn}.

The uncertainties are summarized in Table~\ref{tab:flux-uncertainty}. The dominant systematic uncertainty is induced by the variations in the energy scale, acrylic vessel radius, and CJPL-I's latitude and longitude.

\begin{table}[!htbp]
    \tabcolsep=0.3cm
    \centering
    \caption{Systematic uncertainty of flux measurement}
    \begin{tabular}{l c c}
        \hline
        \hline
        Source                   & Uncertainty & Flux Uncertainty \\
        \hline
        Energy scale             & $\pm5\%$    & $\pm1.6\%$       \\
        Acrylic vessel radius    & $\pm0.5$~cm & $\pm1.5\%$       \\
        Lead shielding thickness & $\pm5$~cm   & $\pm0.2\%$       \\
        Rock thickness           & $\pm50$~cm  & $\pm0.7\%$       \\
        Physical model           & $\pm50\%$   & $\pm0.5\%$       \\
        Muon generator           & ~           & $\pm0.2\%$       \\
        Latitude and longitude   & $\pm100$~m  & $\pm1.1\%$       \\
        Elevation                & $\pm100$~m  & $\pm0.6\%$       \\
        \\
        Total                    & ~           & $\pm2.7\%$       \\
        \hline
        \hline
        \label{tab:flux-uncertainty}
    \end{tabular}
\end{table}

The energy scale uncertainty is determined to be 5\% at the low energy region ($\sim$ 1~MeV). However, the energy scale factor in the unit of PE/MeV is lower for the muon events, due to the waveform saturation and energy non-linearity effect. Therefore, the energy scale for muons in simulation is fine-tuned by a binned $\chi^2$ minimization method to ensure a good consistency with data. The $\chi^2$ is defined as,
\begin{equation}
    \chi^2 = \sum_i \frac{\left[n^i_\mathrm{data}-n^i_\mathrm{sim}(\eta)\right]^2}{n^i_\mathrm{data}},
\end{equation}
where $n^i_\mathrm{data}$ is the number of events in the $i$-th bin of the energy distribution of data, $n^i_\mathrm{sim}(\eta)$ is the number of events in the $i$-th bin of the energy distribution of the MC sample, $\eta$ is the additional energy scale shift applied on an event-by-event basis for the MC sample. The uncertainty in $\eta$ is defined by the range over which adjusting $\eta$ changes $\chi^2$ by $\pm1$ from its minimum value. The fitting provides that $\eta=0.87\pm0.01$, presenting the effects from the energy non-linearity and waveform saturation. This uncertainty, i.e., $\sim1\%$ is covered by the energy scale uncertainty from the low-energy region. Given this fact, and the non-linearity and waveform saturation have not yet been studied comprehensively, the final analysis still adopts the 5\% uncertainty for the energy scale conservatively. The corresponding muon flux uncertainty is estimated to be 1.6\%.

The radius of acrylic vessel constrains the target volume directly affects the deposited energy of muons, and then the estimation of muon detection efficiency $\varepsilon_\mathrm{d}$. In the production of the acrylic vessel, the machining accuracy of radius is about 0.5~cm, which is set as its uncertainty. In simulation geometry, a 0.5~cm variation is implemented for the radius of target volume, resulting in a 1.5\% uncertainty for muon flux.

The thickness of lead shielding and rock can affect the contribution of muon showers. The lead shielding is arranged brick by brick. There should be a tiny difference between the actual detector geometry and that used in \GEANT4 simulation, where the lead shielding is considered as a prefect circular cylindrical shell. Assuming a 5~cm thickness variation conservatively, which is the typical size of a lead brick, it contributes little uncertainty to the overall efficiency.

The rock thickness of the laboratory in the simulation is set to be 1~m and varied by $\pm0.5$~m to validate this arrangement. The variation is found to have little impact on the overall efficiency. The average free path of neutrons at 50~MeV is theoretically calculated to be about 10~cm in the rock around CJPL-I, indicating the rock's thickness is reasonable for the study of the muon shower effect. The uncertainty of secondary particle yield in muon showers is conservatively taken to be 50\% for uncertainty from the physical models to describe the physical processes of muon showers generation. Only a 0.5\% uncertainty to the muon flux is found.

The equivalent projection area of the experiment hall is sensitive to the angular distribution of underground muon. Therefore, the different inputs for underground muon can change the calculation result. Due to the complexity of the tunnel direction and the overburden terrain at CJPL-I, the accurate location of laboratory is hard to be obtained. We varied the location of the laboratory around the nominal one in latitude, longitude and elevation to study the effects. Additionally, the primary muon generators, the modified Gaisser's formula and MCEq, are implemented separately, confirming that the induced difference in the projection area is negligible.

Using the selected muon candidates and considering the systematic uncertainties described above, we measure the cosmic ray flux at CJPL-I to be
\begin{equation}
    \phi_\mu = (3.56\pm0.16_{\mathrm{stat.}}\pm0.10_{\mathrm{syst.}})\times10^{-10}~\mathrm{cm}^{-2}\mathrm{s^{-1}}.
\end{equation}
This measurement is in excellent agreement with the previous one, but the precision is increased by $\sim25\%$.

\section{Measurement of cosmogenic Neutron Yield in liquid scintillator}\label{sec:Neutrons}
Cosmogenic neutrons are primarily generated in hadronic and electromagnetic showers that result from the passage of high-energy cosmic muons~\cite{Li:2015kpa}. The neutron yield induced by muons is defined as the production rate of neutrons per unit muon track length per unit target density~\cite{SNO:2019pzy} and can be expressed as
\begin{equation}
    Y_n = \frac{N_n}{(\sum L_{\mu}) \times \rho} = \frac{N_n}{N_\mu \times L_{avg} \times \rho},
    \label{equ:yield}
\end{equation}
where $N_n$ represents the number of neutrons generated directly or indirectly when muons pass through the target volume, $N_\mu$ denotes the number of muons traversing the target volume, $L_{\mathrm{avg}}$ is the average path length of muons within the target volume, and $\rho$ signifies the density of the material.

\subsection{Event Selection}
In LS, the neutrons produced by muons are dominantly captured by hydrogen ($n$H) after thermalization, emitting a 2.2-MeV $\gamma$. The coincidence between muon and this typical $\gamma$ is used to tag the cosmogenic neutron events in the detector. Events following the tagged muon signals are considered neutron candidates. The time interval $t_n$ between a neutron candidate and its preceding muon event must be within the range of [20, 1020]~$\mu\mathrm{s}$, and the energy of neutron candidate must be within the range of [0, 4]~MeV. The lower bound of $t_n$ excludes after-pulses, re-triggers and significant baseline fluctuations caused by the muon's passage, and cosmogenic nuclei with short lifetimes. The upper bound of $t_n$ ensures the correlation between the muon and the 2.2-MeV $\gamma$ emitted by $n$H capture, typically occurring with a capture time of approximately 200~$\mu\mathrm{s}$. In this study, the number of neutrons candidates $m$ passing the selection criteria is 68.

\subsection{Determining the Signal and Background}
The neutron candidates include lots of accidental coincidences from natural radioactivity. We used an unbinned likelihood fit method with the two-dimension distribution of energy and time interval $t_n$ to determine $N_{\mathrm{obs}}$. Assuming the number of neutron candidates follows a Poisson distribution with an expected value of $\nu$, the likelihood function can be expressed as shown in Eq.~\ref{eq:likelihood}, with the energy and time distributions treated as independent.
\begin{equation}\label{eq:likelihood}
    \mathcal{L} = \frac{\nu^m}{m!}e^{-\nu}\prod^{m}_{i=1}[f_1(E_i, \kappa)f_2(t_i)]N(\kappa),
\end{equation}
where $i$ represents the $i$-th selected events in the data, $f_1(E_i, \kappa)$ represents the energy distribution of neutron candidates, and $f_2(t_i)$ represents the time distribution, $N(\kappa)$ is the zero-centered Gaussian distribution with a 5\% standard deviation, accounting for the uncertainty from energy scale.

The energy distribution is defined as
\begin{equation}
    f_1(E, \kappa) = \omega f_s[E(1+\kappa)] + (1-\omega) f_b[E(1+\kappa)],
\end{equation}
where $\omega$ is the ratio of the signal rate to the total event rate, $f_s$ is the calorimeter function, which can well describe both the complete and partial absorption of gamma energy in the detector~\cite{cheng2016determination}, $f_b$ is the distribution of the backgrounds from accidental coincidences.
The background spectrum is estimated by analyzing events far from the muon events in time. To enhance the background statistics, we set the time window for background selection to $[2000, 12000]~\mathrm{\mu s}$ after the preceding muon. Besides the peak of 2.61-MeV $\gamma$ from $^{208}\mathrm{Tl}$ decay, there are two more peaks in the background distribution around 0.2~MeV and 1.4~MeV originating from the two trigger thresholds in different periods of the detector.

The distribution of time interval $t_n$ is described by
\begin{equation}
    f_2(t) = \omega A e^{-t_n/{\tau_n}} + (1-\omega) B,
\end{equation}
where $\tau_n$ is the mean $n$H capture time, $A$ and $B$ are the normalization factors for exponential and uniform distributions in the selection time interval, respectively.

\begin{figure}[!h]
    \centering
    \subfloat[]{
        \includegraphics[width=0.95\columnwidth]{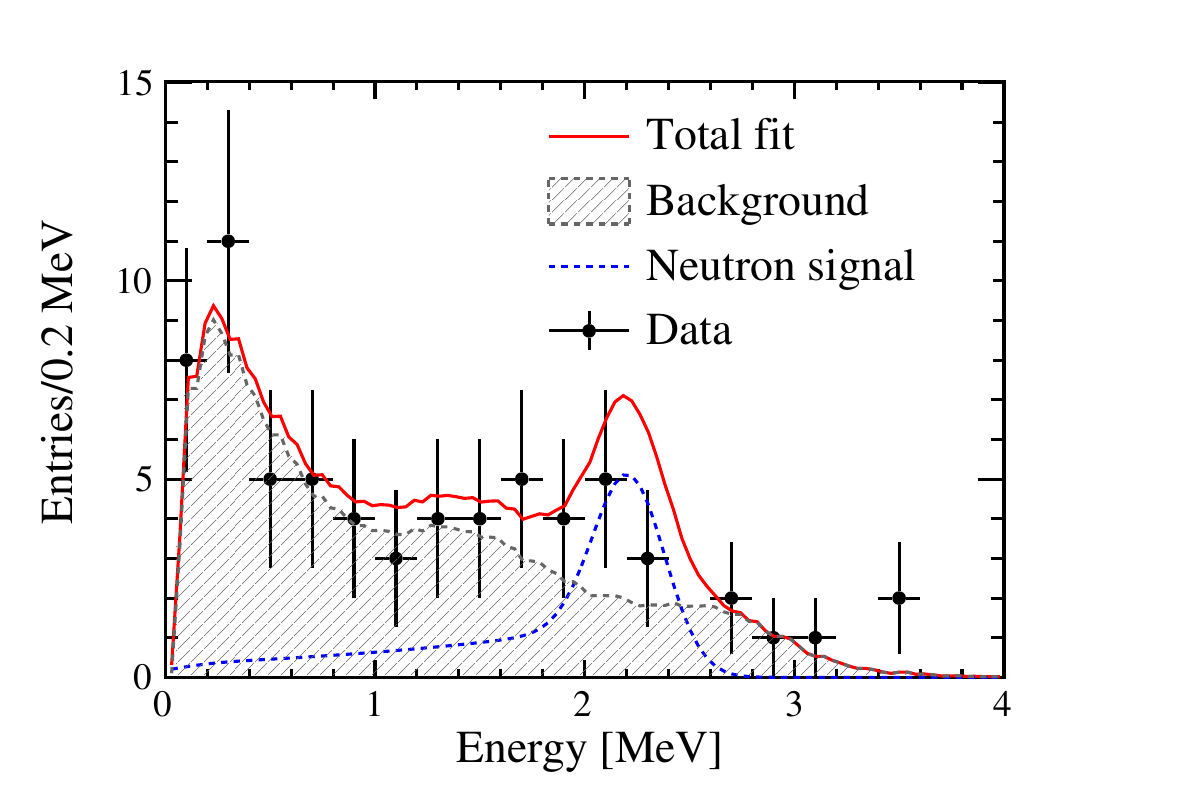}
    }\hspace*{\fill}\\
    \subfloat[]{
        \includegraphics[width=0.95\columnwidth]{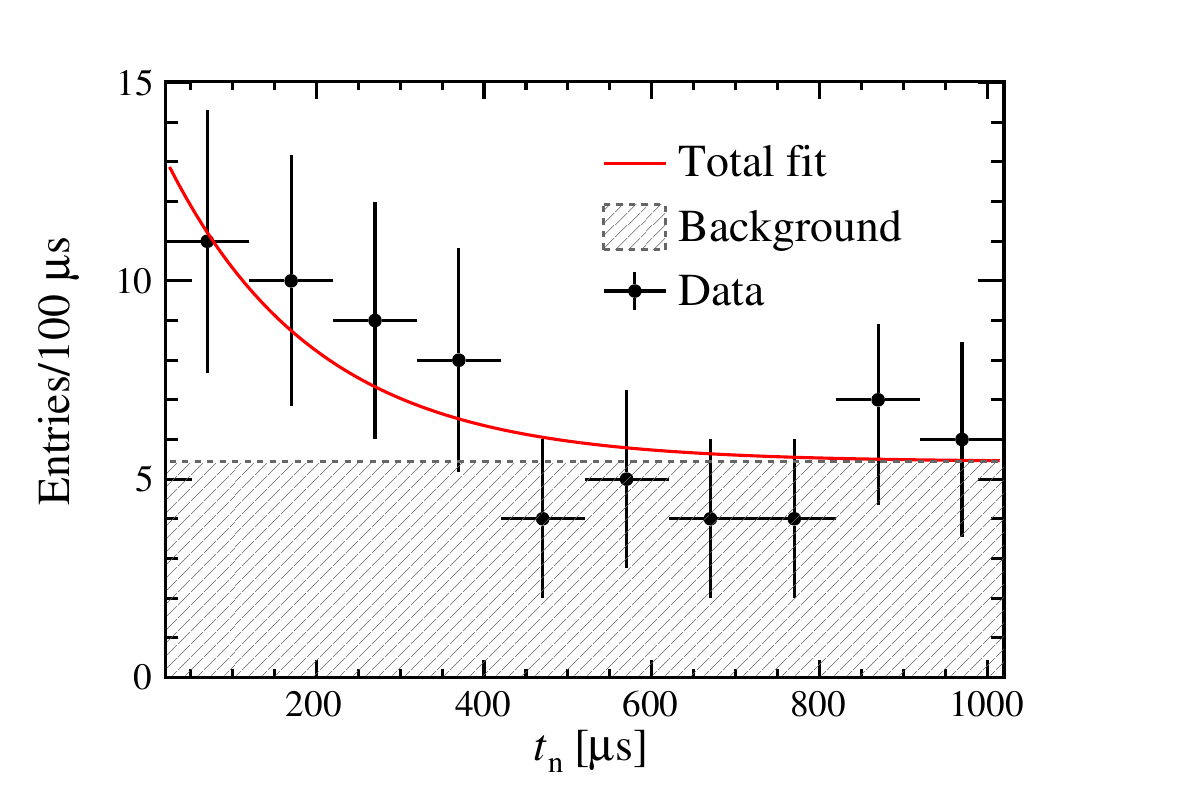}
    }
    \caption{The energy and time interval spectra from the two-dimension likelihood fit result. The data events are selected within $[20, 1020]~\mu\mathrm{s}$ time interval from preceding muons. The gray area represents the background distribution estimated by the events far from the muon signals. The dashed line shows the $n$H capture signals described by the calorimeter function~\cite{cheng2016determination}, while the solid line shows the total fit result.}
    \label{fig:energy}
\end{figure}

Fig.~\ref{fig:energy} shows the projected energy and time distributions obtained from the two-dimension fit. The $\chi^2/\mathrm{ndf}$ of this fit is calculated to be $12.3/18$ under the binning shown in the figure. $N_{\mathrm{obs}}$ is determined to be $\omega \times m = 13.6\pm5.7$, where the uncertainty is dominated by the statistics of data. We also changed the time range for background event selection to study the uncertainty from the function $f_b$ and the result indicates that it is negligible.

To validate the fit result, a counting method is also used to subtract the background from the neutron candidates. We apply a normalization factor $k$ to estimate the background counts in signal's time window by scaling events from background's time window~\cite{DayaBay:2017txw}. In this analysis, due to the low rate of muon events, this factor is simplified to the ratio of the interval lengths. The number of neutron signal events is given by
\begin{equation}
    N_{\mathrm{obs}} = N_c - k \times N_b,
\end{equation}
where $N_c$ represents the number of neutron candidates, $N_b$ represents the number of background events.

For this method, the events within $[1.5, 2.9]$~MeV, approximating the $\pm3\sigma$ energy region of 2.2-MeV $\gamma$ observed in the detector, are used. By counting the event number in different time window, $N_c$ is $23.0\pm4.8$ while $N_b$ is $121.0\pm11.0$ in the data. As the normalization factor $k$ is calculated as 0.1, $10.5\pm4.9$ neutron signals are obtained, where the dominant uncertainty for $N_{\mathrm{obs}}$ is from the statistics of $N_{\mathrm{c}}$, which contributes 99.7\% in total. The results from these two methods are consistent, but the fit method gives a better precision.

\subsection{Efficiencies, Correction Factors and Uncertainties}\label{sec:neutron-eff}
In addition to the number of candidates $N_{\mathrm{obs}}$ from event selection and signal fit, further efficiencies and corrections are needed to calculate $N_\mathrm{n}$, including the neutron detection efficiency $\varepsilon_\mathrm{nt}$, time selection efficiency $\varepsilon_\mathrm{t}$, and also the non-target factor for neutron capture on non-target volumes $f_\mathrm{non-target}$, shower factor $f_\mathrm{shower}$, neutron spill factor $f_\mathrm{spill}$, detector's finite-size factor $f_\mathrm{size}$. The overall relationship can be written as
\begin{equation}
    N_\mathrm{n} = N_\mathrm{obs}\times \frac{(1-f_\mathrm{non-target}) \times (1-f_\mathrm{shower}) \times f_\mathrm{spill} }{\varepsilon_\mathrm{nt} \times \varepsilon_\mathrm{t}\times f_\mathrm{size}}.
\end{equation}
Additionally, parameters such as muons' average path length and the density of LS also significantly impact the measurement results of cosmogenic neutron yield.

This section focuses on evaluating the nominal values and uncertainties of these parameters. Systematic uncertainties are examined by comparing the variations in values for different detector geometries, muon generators, and physical models used in the simulations. The energy scale uncertainty has already been accounted for in the fitted signal number by incorporating it into Eq.~\ref{eq:likelihood}.

In this study, the density of LS is measured to be 0.86~$\mathrm{g}/\mathrm{cm}^3$ and the average path length $L_{\mathrm{avg}}$ of muons is derived from the detector simulations, as described in Sec.~\ref{sec:Simulation}. Following the application of the muon selection criteria detailed in Sec.~\ref{sec:Muons}, $L_{\mathrm{avg}}$ in the LS is calculated to be 95.6~cm with negligible statistical uncertainty. A variation of 0.5~cm in the acrylic vessel radius can alter the target volume of LS, resulting in a 1.4\% uncertainty in $L_{\mathrm{avg}}$. The energy scale, which influences the selection of muons based on their energy, can affect the muon candidates, and subsequently, $L_{\mathrm{avg}}$. This impact is assessed using the MC sample, revealing that a 5\% uncertainty in the energy scale introduces a 0.6\% uncertainty in $L_{\mathrm{avg}}$. Additionally, variations between primary muon generators can lead to a 0.3\% uncertainty due to the slight differences of angular distribution. Finally, the total uncertainty in $L_{\mathrm{avg}}$ is determined to be 1.5\%.

The neutron trigger efficiency, $\varepsilon_\mathrm{nt}$, is derived from the MC sample to account for the trigger efficiency of 2.2-MeV $\gamma$ from $n$H captures. Due to the varying operational states of the detector, as outlined in Table~\ref{tab:trigger-conditions}, $\varepsilon_\mathrm{nt}$ exhibits slight variations across different periods. We calculated $\varepsilon_\mathrm{nt}$ for each period, and the weighted average over the effective DAQ time is used for the final measurement, numerically 63.8\%. The dominant uncertainty of $\varepsilon_\mathrm{nt}$ is from the variation in the target volume's radius, and evaluated to be 1.4\%.

The time selection efficiency, $\varepsilon_\mathrm{t}$, is also derived from the simulations as 90.6\% to account for the impact of time selection. The time interval between the muon and neutron events is determined by the neutron generation time $t_g$ and the neutron capture time $t_c$. Only $t_g$ varies slightly for the three physical models used in the simulation, with differences on the order of sub-nanoseconds, which are very short compared to $t_c$. Consequently, the total uncertainty in $\varepsilon_\mathrm{t}$ is evaluated as 0.8\% which is from the MC statistics.
\subsubsection{Non-Target Correction}
Neutrons produced by tagged muons in the non-LS region within the detector can be captured in the non-LS region. The main contribution of this effect is from the water and acrylic materials. The gamma rays emitted in this process may enter the LS and generate a trigger passing the selection criteria, mimicking the neutron signal. To account for this effect, we define a non-target correction factor, $f_\mathrm{non-target}$, as the ratio of gamma rays entering the LS, which are emitted by neutrons generated and captured outside the LS, to the total number of gamma candidates. The MC sample indicates that the nominal value of $f_\mathrm{non-target}$ is 0.24.
\subsubsection{Shower Correction}
Muons can produce showers in the water, lead shielding, and rock, generating secondary particles with high kinetic energy, such as fast neutrons. Although these muons do not traverse the LS, some events are still incorrectly tagged as muon signals due to the muon shower, and the subsequent neutron captures are included in the data sample. The factor $f_\mathrm{shower}$ is used to correct this effect and is evaluated in the MC sample as the ratio of the number of neutron candidates following the muon shower events to those following muon showers or traversing muons. In the simulation, approximately 2.0\% of tagged muon events are associated with muon showers, resulting in the shower correction factor as 0.21. $80\%$ of the contribution to this correction factor is from the high-energy neutrons, the other $20\%$ is from gammas.
\subsubsection{Neutron Spill Correction}
The neutron spill factor $f_\mathrm{spill}$ is applied to account for the net effect of spill-in and spill-out of neutrons induced by muons. Neutrons produced inside the LS by tagged muons can propagate to escape the LS before being captured and are treated as spill-out neutrons. In contrast, neutrons produced outside the LS can propagate to be captured inside the LS and be regarded as spill-in neutrons. Additionally, the secondary high-energy particles induced by muons outside the LS can spill into the LS and produce neutrons. This process can also contribute 47.0\% of the spill-in neutrons. The correction factor $f_\mathrm{spill}$ is defined as the ratio of the number of neutrons produced in the LS due to the muon's passage, to the number of neutrons captured in the LS. The nominal value of this factor is derived from simulation as 0.61.
\subsubsection{Uncertainties of $f_\mathrm{non-target}$, $f_\mathrm{shower}$, and $f_\mathrm{spill}$}
The values of correction factor $f_\mathrm{non-target}$, $f_\mathrm{shower}$, and $f_\mathrm{spill}$ could differ with various combinations of neutron yields in LS and other materials. The predicted neutron yields in materials vary among the three physical models in the simulation, providing an estimation of the systematic uncertainties for these three factors. Each factor is derived from three MC samples using these three physical models, in which the differences in the values are considered as systematic uncertainties. The uncertainties from MC statistics are subdominant but are still included in the final results. For $f_\mathrm{non-target}$, an 8.5\% uncertainty comes from physical models, and 3.5\% is from MC statistics. For $f_\mathrm{shower}$, 9.7\% of the uncertainty is addressed by comparing results between different physical models, while 1.4\% comes from MC statistics. For $f_\mathrm{spill}$, a 6.4\% uncertainty is induced by differences in physical models, and 4.7\% originates from MC statistics. In conclusion, the total uncertainties considered for $f_\mathrm{non-target}$, $f_\mathrm{shower}$ and $f_\mathrm{spill}$, are 9.2\%, 9.8\%, 7.9\% respectively.

\subsubsection{Finite-Size Effect}
For the production of neutrons in the LS induced by muons reaching CJPL-I, the direct muon spallation with carbon nuclei only contributes about 3.8\% in simulation, while the dominant parts come from the indirect processes initiated by muon-induced secondary particles, such as high-energy gammas. Due to the finite size of the detector, secondary particles produced by muons in the LS can escape from the target volume, LS. If the detector size were slightly enlarged, these secondary particles might generate additional neutrons, which should be included when evaluating the neutron yield. While this additional neutron production contributes negligibly to the neutron yield in large detectors, it can be visible in small-size detectors, like in the 1-ton prototype. To ensure a consistent definition of neutron yield with other experiments and to make the measured yield independent of detector size, this effect is studied in detail and then is used for correction in the neutron yield measurement.

We construct a sphere with an 8-meter radius filled with LS in \GEANT4 to study this effect. The volume within a sphere of the 0.645-meter radius is considered as the target region of the 1-ton prototype, while the volume outside is treated as the buffer region. The tracks of secondary particles produced inside the target region due to the passage of muons are recorded. When counting neutrons, the $(n,2n)$ reaction increases the neutron number by 1, while the $(n,n^\prime)$ reaction does not change the neutron number. To correct for the finite-size effect described above, we calculate the factor $f_\mathrm{size}$ as follows:
\begin{equation}
    f_\mathrm{size}=\frac{N_\mathrm{in}}{N_\mathrm{in}+N_\mathrm{out}},
\end{equation}
where $N_\mathrm{in}$ is the number of neutrons produced by muons inside the target region, either directly or indirectly, and $N_\mathrm{out}$ is the number of neutrons generated in the buffer region by secondary particles of muons emitted from the target region.

\begin{figure}[!htbp]
    \centering
    \includegraphics[width=0.95\columnwidth]{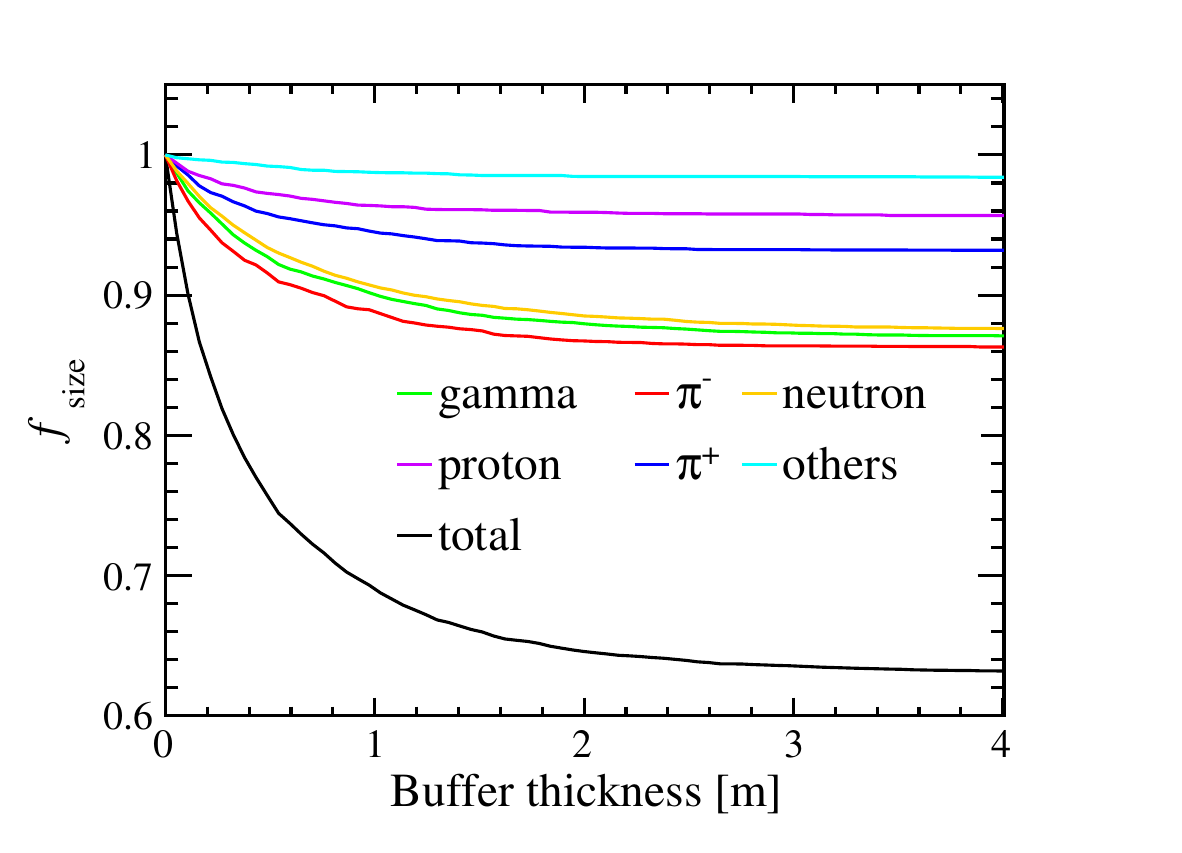}
    \caption{The finite-size correction factors $f_\mathrm{size}$ are evaluated with different buffer thickness in the simulation. The contributions from different particles to this factor are also plotted.}
    \label{fig:factor}
\end{figure}

The factors $f_\mathrm{size}$ are evaluated with different buffer thickness using the MC sample and shown in Fig.~\ref{fig:factor}. The results indicate that a buffer region with at least a 3-meter thickness is required for the detector and the corresponding correction factor is 0.63, which is taken as the nominal value of the factor $f_\mathrm{size}$ in the following measurement. The contributions of different secondary particles are also depicted in Fig.~\ref{fig:factor}. The inelastic scatterings of muon-induced secondary high-energy particles on carbon nuclei are the dominant sources of this effect, including protons, neutrons, and positive and negative pions ($\pi^+$ and $\pi^-$). For $\pi^-$, thermal captures on carbon and hydrogen nuclei, such as the $\pi^-(p,n)\pi^0$ reaction, also contribute to this effect. Photonuclear disintegration on carbon nuclei, typically the $\gamma(^{12}\mathrm{C},^{11}\mathrm{C})n$ reaction, is another significant source. Other high-energy particles like electrons, positrons, deuterons, tritons, and strange hadrons also make few contributions through inelastic scatterings. Additionally, muons can generate radioactive nuclei in the LS, such as $^{4}\mathrm{H}$, which can produce neutrons through decay. However, these nuclei are unlikely to spill out before decaying due to their heavy mass, thus contributing minimally to the correction factor. The particle composition accounting for this effect is similar to that in previous simulation studies for cosmogenic neutrons~\cite{Wang:2001fq}.

The yields of secondary particles produced by muons and the neutron yield of secondary particles in LS could differ among the three physical models. Therefore, the factor $f_\mathrm{size}$ is studied using each of the three models, with numerical differences amounting to 1.4\% covered by the statistical uncertainty of 1.3\%. A 5~mm variation in the detector radius, which corresponds to the parameter uncertainty of the acrylic vessel radius, can induce a 0.3\% uncertainty in $f_\mathrm{size}$. The total uncertainty of $f_\mathrm{size}$ is thus estimated to be 1.3\%, calculated as the quadratic sum of the two components.

\begin{figure}[!htbp]
    \centering
    \includegraphics[width=0.95\columnwidth]{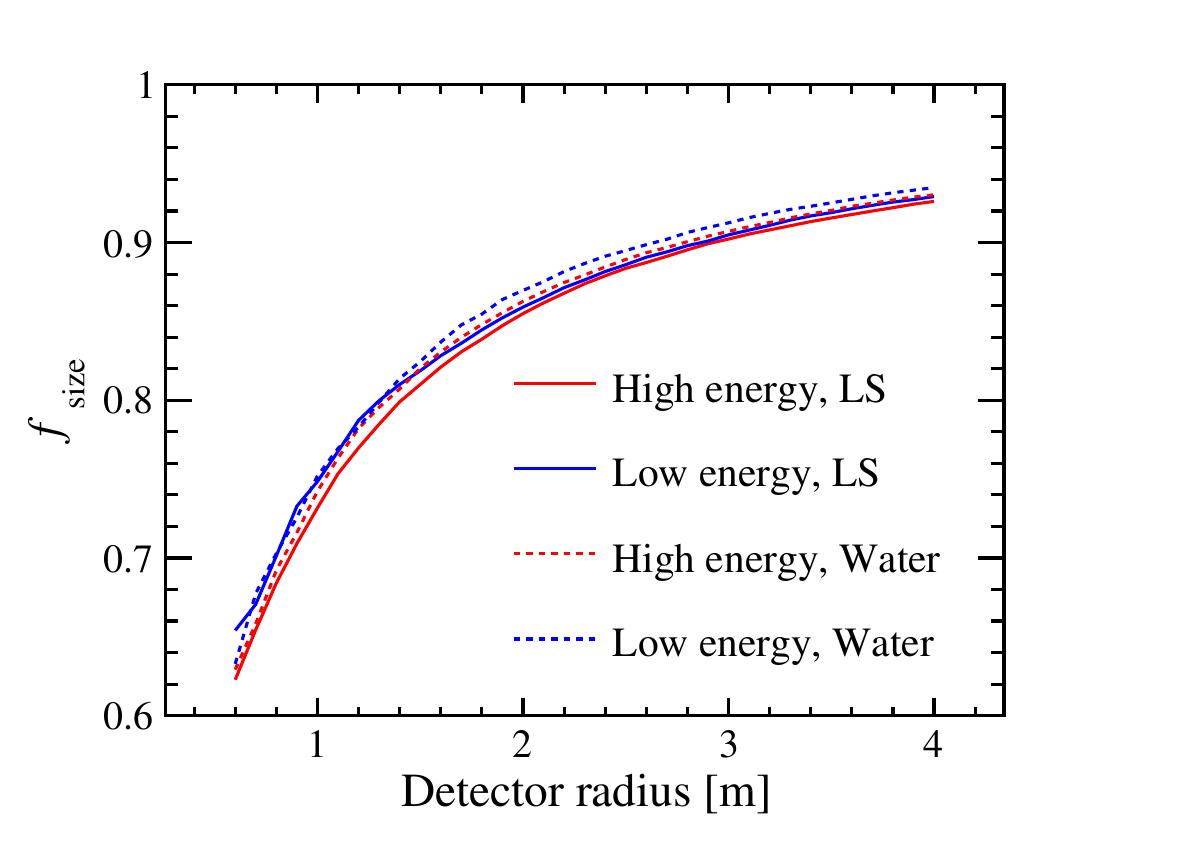}
    \caption{The finite-size correction factor $f_\mathrm{size}$ varies with the detector size, while the buffer thickness outside the detector is fixed at 3~m. The red lines represent the muons with $\sim360$~GeV average energy corresponding to the muons arriving at CJPL-I. The blue lines show the low-energy muons with 70~GeV average energy. The correction factor in LS- and water-based detectors are plotted with solid and dashed lines, respectively.}
    \label{fig:size}
\end{figure}

Furthermore, the radius of the target region is also varied with a fixed buffer thickness of 3~meters to study the relationship between the correction factor and detector size. The result, shown in Fig.~\ref{fig:size}, demonstrates that this effect is sensitive to detector size. Additionally, we also studied the impact of different average muon energies and target materials, which are the dominant determinant for this factor. For experiments with shallow overburden such as the experimental hall-1(EH1) of Daya Bay, where the average muon energy is about 70~GeV~\cite{DayaBay:2017txw}, this effect is slightly weakened, primarily due to the increased ratio of neutrons produced through direct inelastic scattering by low-energy muons~\cite{Wang:2001fq}. This effect is also weaker in water than in LS due to secondary particles' shorter average free path. For example, in simulations, the average free path of secondary gamma rays produced by muons in water is calculated to be 46.6~cm, while it becomes 66.5~cm in LS, resulting in fewer spill-out of gammas.

This effect could also be significant for measuring other cosmogenic production yields in small-size detectors, such as for $^{9}\mathrm{Li}$ and $^{8}\mathrm{He}$. These radioactive isotopes are predominantly produced by secondary particles like $\pi^-$ induced by muons~\cite{Super-Kamiokande:2015xra, Li:2015kpa}.

Table~\ref{tab:yield uncertainty} summarizes the systematic uncertainties of correction factors, efficiencies, and other parameters for this measurement.
\begin{table}[!htbp]
    \tabcolsep=0.22cm
    \centering
    \caption{The contributions of systematic uncertainties in the measurement of cosmogenic neutron yield, in which the dominant source is also stated.}
    \begin{tabular}{l    c   l}
        \hline
        \hline
        Parameter                & Uncertainty Contribution & Dominant source       \\
        \hline
        $\varepsilon_\mathrm{nt
        }$                       & $\pm1.4\%$               & Acrylic vessel radius \\
        $\varepsilon_\mathrm{t}$ & $\pm0.8\%$               & MC statistic          \\
        $f_\mathrm{non-target}$  & $\pm2.9\%$               & Physical model        \\
        $f_\mathrm{shower}$      & $\pm2.6\%$               & Physical model        \\
        $f_\mathrm{spill}$       & $\pm7.9\%$               & Physical model        \\
        $f_\mathrm{size}$        & $\pm1.9\%$               & MC statistic          \\
        $L_\mathrm{avg}$         & $\pm1.3\%$               & Acrylic vessel radius \\
        \\
        Total                    & $\pm9.3\%$               &                       \\
        \hline
        \hline
    \end{tabular}

    \label{tab:yield uncertainty}
\end{table}

\subsection{Result and Comparison}
Using Eq.~\ref{equ:yield}, the cosmogenic neutron yield in LS at CJPL-I is measured to be
\begin{equation}
    Y_\mathrm{n} = (3.37\pm 1.41_{\mathrm{stat.}}\pm 0.31_{\mathrm{syst.}}) \times 10^{-4}~\mathrm{\mu}^{-1} \mathrm{g}^{-1} \mathrm{cm}^{2}.
\end{equation}
The corresponding average muon energy is derived from simulation as $360\pm10$~GeV, which indicates that this work provides the world's highest average muon energy measurement of cosmogenic neutron yield in LS. The uncertainty of average muon energy mainly comes from the differences between two muon generators, MCEq and the modified Gaisser's formula, while the laboratory elevation uncertainty also has subdominant contributions.

Three methods are used to give a prediction for the neutron yield, which can be compared with the measurement result. Firstly, the detector simulations in \GEANT4 provide the predicted value of different models and only the MC statistic uncertainties are considered. Secondly, the result of \Fluka-based simulation study from Ref.~\cite{Wang:2001fq} is also used to provide the predication where only the uncertainty of muon energy is considered. At last, the relation between average muon energy and neutron yield can be described by an empirical formula from previous studies~\cite{Mei:2005gm}, written as
\begin{equation}
    Y_\mathrm{n} = a \times {E_\mathrm{\mu}}^b,
    \label{eq:extrapolation}
\end{equation}
where $a$ and $b$ are parameters under determination. Using the previous measurements from Hertenberger~\cite{Hertenberger:1995ae}, Bohem~\cite{Boehm:2000ru}, Daya Bay (DYB)~\cite{DayaBay:2017txw}, Aberdeen Tunnel~\cite{AberdeenTunnelExperiment:2015uaa}, KamLAND~\cite{KamLAND:2009zwo}, LVD~\cite{LVD:1999ezh} with corrections~\cite{Mei:2005gm}, Borexino~\cite{Borexino:2013cke}, we fit these parameters at low muon energy, considering the measurement uncertainties in each experiment. Then, the neutron yield at $360\pm10$~GeV is predicted by extrapolating Eq.~\ref{eq:extrapolation} with these fitting parameters. Table~\ref{tab:yield predication} shows the predicted values of neutron yield from these methods, which are consistent with the measurement of this work.

\begin{table}[!htbp]
    \tabcolsep=0.55cm
    \centering
    \caption{The predicted cosmogenic neutron yield with average muon energy of $360$~GeV in simulations and the extrapolation from other experimental measurements.}
    \label{tab:yield predication}
    \begin{tabular}{c  c}
        \hline
        \hline
        Method              & Yield~[$10^{-4}\mathrm{\mu}^{-1}\mathrm{g}^{-1}\mathrm{cm}^{2}$] \\
        \hline
        \GEANT4 (QGSP-BERT) & $3.65\pm0.04$                                                    \\
        \GEANT4 (QGSP-BIC)  & $3.13\pm0.03$                                                    \\
        \GEANT4 (FTFP-BERT) & $3.56\pm0.04$                                                    \\
        \Fluka              & $ 3.23\pm0.07$                                                   \\
        Extrapolation       & $3.63\pm1.59$                                                    \\

        \hline
        \hline
    \end{tabular}
\end{table}

\begin{figure}[!htbp]
    \centering
    \includegraphics[width=0.95\columnwidth]{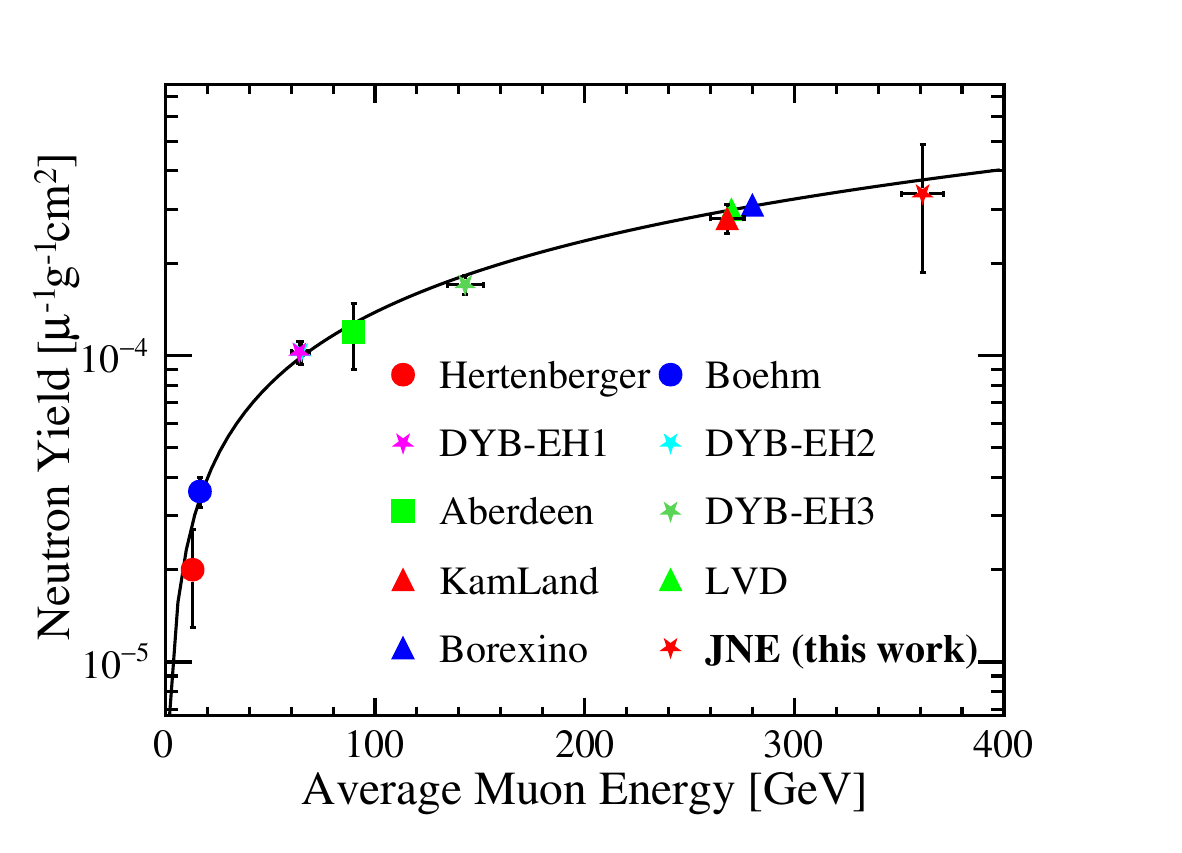}
    \caption{The cosmogenic neutron yield as a function of average muon energy from this study is compared with results from other experiments. The solid line represents the global fit obtained using the power-law formula.}
    \label{fig:yield}
\end{figure}

The cosmogenic neutron yields of different LS detectors in terms of average muon energy are shown in Fig.~\ref{fig:yield}. Combined with the measurement in this work, the global fit using Eq.~\ref{eq:extrapolation} is processed. The fit result shows that
\begin{equation}
    a = (3.9\pm0.7)\times 10^{-6}~\mathrm{\mu}^{-1} \mathrm{g}^{-1} \mathrm{cm}^{2},
    b = 0.77\pm0.03,
\end{equation}
which is consistent with previous study within the range of uncertainty~\cite{DayaBay:2017txw}.

\section{Summary and discussion}\label{sec:Summary}
In this work, we study the cosmic muons reaching CJPL-I using 1178 days of data collected by the 1-ton prototype neutrino detector for the Jinping Neutrino Experiment. Comprehensive simulations are also conducted to understand the interaction processes of muons with mountain rocks and the detector. Through a thorough evaluation of detection efficiency and systematic uncertainties, this study determines the cosmic muon flux to be $(3.56\pm0.16_{\mathrm{stat.}}\pm0.10_{\mathrm{syst.}})\times10^{-10}~\mathrm{cm}^{-2}\mathrm{s^{-1}}$ at CJPL-I, and the cosmogenic neutron yield in liquid scintillator to be $(3.37\pm 1.41_{\mathrm{stat.}}\pm 0.31_{\mathrm{syst.}}) \times 10^{-4}~\mathrm{\mu}^{-1} \mathrm{g}^{-1} \mathrm{cm}^{2}$ with about $360$~GeV average muon energy. The reconstructed angular distribution of muons is in good agreement with our simulations that take into account the mountain's topography near CJPL-I.

This study successfully addresses the challenges of measuring the cosmogenic production yield in a small-size LS detector. The proposed finite-size correction factor is investigated in detail under different buffer thickness, muons' average energy, target materials and detector's size. The information should be valuable for the related study in other experiments.  A measurement of cosmogenic neutron yield for LS is presented with the highest average muon energy in the world. The measurement results will serve as important references for conducting low-background experiments in the future at CJPL.

\section{Acknowledgements}
This work was supported in part by the National Natural Science Foundation of China (12127808, 12141503 and 12305117) and the Key Laboratory of Particle and Radiation Imaging (Tsinghua University). We acknowledge Orrin Science Technology, Jingyifan Co., Ltd, and Donchamp Acrylic Co., Ltd, for their efforts in the engineering design and fabrication of the stainless steel and acrylic vessels. Many thanks to the CJPL administration and the Yalong River Hydropower Development Co., Ltd. for logistics and support.

\bibliographystyle{apsrev4-2}
\bibliography{bibfile}

\begin{thebibliography}{43}%
\makeatletter
\providecommand \@ifxundefined [1]{%
 \@ifx{#1\undefined}
}%
\providecommand \@ifnum [1]{%
 \ifnum #1\expandafter \@firstoftwo
 \else \expandafter \@secondoftwo
 \fi
}%
\providecommand \@ifx [1]{%
 \ifx #1\expandafter \@firstoftwo
 \else \expandafter \@secondoftwo
 \fi
}%
\providecommand \natexlab [1]{#1}%
\providecommand \enquote  [1]{``#1''}%
\providecommand \bibnamefont  [1]{#1}%
\providecommand \bibfnamefont [1]{#1}%
\providecommand \citenamefont [1]{#1}%
\providecommand \href@noop [0]{\@secondoftwo}%
\providecommand \href [0]{\begingroup \@sanitize@url \@href}%
\providecommand \@href[1]{\@@startlink{#1}\@@href}%
\providecommand \@@href[1]{\endgroup#1\@@endlink}%
\providecommand \@sanitize@url [0]{\catcode `\\12\catcode `\$12\catcode `\&12\catcode `\#12\catcode `\^12\catcode `\_12\catcode `\%12\relax}%
\providecommand \@@startlink[1]{}%
\providecommand \@@endlink[0]{}%
\providecommand \url  [0]{\begingroup\@sanitize@url \@url }%
\providecommand \@url [1]{\endgroup\@href {#1}{\urlprefix }}%
\providecommand \urlprefix  [0]{URL }%
\providecommand \Eprint [0]{\href }%
\providecommand \doibase [0]{https://doi.org/}%
\providecommand \selectlanguage [0]{\@gobble}%
\providecommand \bibinfo  [0]{\@secondoftwo}%
\providecommand \bibfield  [0]{\@secondoftwo}%
\providecommand \translation [1]{[#1]}%
\providecommand \BibitemOpen [0]{}%
\providecommand \bibitemStop [0]{}%
\providecommand \bibitemNoStop [0]{.\EOS\space}%
\providecommand \EOS [0]{\spacefactor3000\relax}%
\providecommand \BibitemShut  [1]{\csname bibitem#1\endcsname}%
\let\auto@bib@innerbib\@empty
\bibitem [{\citenamefont {Cheng}\ \emph {et~al.}(2017)\citenamefont {Cheng} \emph {et~al.}}]{Cheng:2017usi}%
  \BibitemOpen
  \bibfield  {author} {\bibinfo {author} {\bibfnamefont {J.-P.}\ \bibnamefont {Cheng}} \emph {et~al.},\ }\href {https://doi.org/10.1146/annurev-nucl-102115-044842} {\bibfield  {journal} {\bibinfo  {journal} {Ann. Rev. Nucl. Part. Sci.}\ }\textbf {\bibinfo {volume} {67}},\ \bibinfo {pages} {231} (\bibinfo {year} {2017})},\ \Eprint {https://arxiv.org/abs/1801.00587} {arXiv:1801.00587 [hep-ex]} \BibitemShut {NoStop}%
\bibitem [{\citenamefont {Zeng}\ \emph {et~al.}(2020)\citenamefont {Zeng}, \citenamefont {Ma}, \citenamefont {Zeng}, \citenamefont {Zeng}, \citenamefont {Yue}, \citenamefont {Cheng},\ and\ \citenamefont {Li}}]{Zeng:2020cyw}%
  \BibitemOpen
  \bibfield  {author} {\bibinfo {author} {\bibfnamefont {W.-H.}\ \bibnamefont {Zeng}}, \bibinfo {author} {\bibfnamefont {H.}~\bibnamefont {Ma}}, \bibinfo {author} {\bibfnamefont {M.}~\bibnamefont {Zeng}}, \bibinfo {author} {\bibfnamefont {Z.}~\bibnamefont {Zeng}}, \bibinfo {author} {\bibfnamefont {Q.}~\bibnamefont {Yue}}, \bibinfo {author} {\bibfnamefont {J.-P.}\ \bibnamefont {Cheng}},\ and\ \bibinfo {author} {\bibfnamefont {J.-L.}\ \bibnamefont {Li}},\ }\href {https://doi.org/10.1007/s41365-020-00760-3} {\bibfield  {journal} {\bibinfo  {journal} {Nucl. Sci. Tech.}\ }\textbf {\bibinfo {volume} {31}},\ \bibinfo {pages} {50} (\bibinfo {year} {2020})}\BibitemShut {NoStop}%
\bibitem [{\citenamefont {Su}\ \emph {et~al.}(2012)\citenamefont {Su}, \citenamefont {Zeng}, \citenamefont {Liu}, \citenamefont {Yue}, \citenamefont {Ma},\ and\ \citenamefont {Cheng}}]{su2012}%
  \BibitemOpen
  \bibfield  {author} {\bibinfo {author} {\bibfnamefont {J.}~\bibnamefont {Su}}, \bibinfo {author} {\bibfnamefont {Z.}~\bibnamefont {Zeng}}, \bibinfo {author} {\bibfnamefont {Y.}~\bibnamefont {Liu}}, \bibinfo {author} {\bibfnamefont {Q.}~\bibnamefont {Yue}}, \bibinfo {author} {\bibfnamefont {H.}~\bibnamefont {Ma}},\ and\ \bibinfo {author} {\bibfnamefont {J.-P.}\ \bibnamefont {Cheng}},\ }\href {https://doi.org/10.3788/HPLPB20122412.3015} {\bibfield  {journal} {\bibinfo  {journal} {High Power Laser and Particle Beams}\ }\textbf {\bibinfo {volume} {24}},\ \bibinfo {pages} {3015} (\bibinfo {year} {2012})}\BibitemShut {NoStop}%
\bibitem [{\citenamefont {Beacom}\ \emph {et~al.}(2017)\citenamefont {Beacom} \emph {et~al.}}]{Jinping:2016iiq}%
  \BibitemOpen
  \bibfield  {author} {\bibinfo {author} {\bibfnamefont {J.~F.}\ \bibnamefont {Beacom}} \emph {et~al.} (\bibinfo {collaboration} {JNE}),\ }\href {https://doi.org/10.1088/1674-1137/41/2/023002} {\bibfield  {journal} {\bibinfo  {journal} {Chin. Phys. C}\ }\textbf {\bibinfo {volume} {41}},\ \bibinfo {pages} {023002} (\bibinfo {year} {2017})},\ \Eprint {https://arxiv.org/abs/1602.01733} {arXiv:1602.01733 [physics.ins-det]} \BibitemShut {NoStop}%
\bibitem [{\citenamefont {Galbiati}\ \emph {et~al.}(2005)\citenamefont {Galbiati}, \citenamefont {Pocar}, \citenamefont {Franco}, \citenamefont {Ianni}, \citenamefont {Cadonati},\ and\ \citenamefont {Schonert}}]{Galbiati:2004wx}%
  \BibitemOpen
  \bibfield  {author} {\bibinfo {author} {\bibfnamefont {C.}~\bibnamefont {Galbiati}}, \bibinfo {author} {\bibfnamefont {A.}~\bibnamefont {Pocar}}, \bibinfo {author} {\bibfnamefont {D.}~\bibnamefont {Franco}}, \bibinfo {author} {\bibfnamefont {A.}~\bibnamefont {Ianni}}, \bibinfo {author} {\bibfnamefont {L.}~\bibnamefont {Cadonati}},\ and\ \bibinfo {author} {\bibfnamefont {S.}~\bibnamefont {Schonert}},\ }\href {https://doi.org/10.1103/PhysRevC.71.055805} {\bibfield  {journal} {\bibinfo  {journal} {Phys. Rev. C}\ }\textbf {\bibinfo {volume} {71}},\ \bibinfo {pages} {055805} (\bibinfo {year} {2005})},\ \Eprint {https://arxiv.org/abs/hep-ph/0411002} {arXiv:hep-ph/0411002} \BibitemShut {NoStop}%
\bibitem [{\citenamefont {Bellini}\ \emph {et~al.}(2014)\citenamefont {Bellini} \emph {et~al.}}]{Borexino:2013zhu}%
  \BibitemOpen
  \bibfield  {author} {\bibinfo {author} {\bibfnamefont {G.}~\bibnamefont {Bellini}} \emph {et~al.} (\bibinfo {collaboration} {Borexino}),\ }\href {https://doi.org/10.1103/PhysRevD.89.112007} {\bibfield  {journal} {\bibinfo  {journal} {Phys. Rev. D}\ }\textbf {\bibinfo {volume} {89}},\ \bibinfo {pages} {112007} (\bibinfo {year} {2014})},\ \Eprint {https://arxiv.org/abs/1308.0443} {arXiv:1308.0443 [hep-ex]} \BibitemShut {NoStop}%
\bibitem [{\citenamefont {Agostini}\ \emph {et~al.}(2021)\citenamefont {Agostini} \emph {et~al.}}]{Borexino:2021pyz}%
  \BibitemOpen
  \bibfield  {author} {\bibinfo {author} {\bibfnamefont {M.}~\bibnamefont {Agostini}} \emph {et~al.} (\bibinfo {collaboration} {Borexino}),\ }\href {https://doi.org/10.1140/epjc/s10052-021-09799-x} {\bibfield  {journal} {\bibinfo  {journal} {Eur. Phys. J. C}\ }\textbf {\bibinfo {volume} {81}},\ \bibinfo {pages} {1075} (\bibinfo {year} {2021})},\ \Eprint {https://arxiv.org/abs/2106.10973} {arXiv:2106.10973 [physics.ins-det]} \BibitemShut {NoStop}%
\bibitem [{\citenamefont {Agostini}\ \emph {et~al.}(2020{\natexlab{a}})\citenamefont {Agostini} \emph {et~al.}}]{BOREXINO:2020hox}%
  \BibitemOpen
  \bibfield  {author} {\bibinfo {author} {\bibfnamefont {M.}~\bibnamefont {Agostini}} \emph {et~al.} (\bibinfo {collaboration} {Borexino}),\ }\href {https://doi.org/10.1140/epjc/s10052-020-08534-2} {\bibfield  {journal} {\bibinfo  {journal} {Eur. Phys. J. C}\ }\textbf {\bibinfo {volume} {80}},\ \bibinfo {pages} {1091} (\bibinfo {year} {2020}{\natexlab{a}})},\ \Eprint {https://arxiv.org/abs/2005.12829} {arXiv:2005.12829 [hep-ex]} \BibitemShut {NoStop}%
\bibitem [{\citenamefont {Abe}\ \emph {et~al.}(2024)\citenamefont {Abe} \emph {et~al.}}]{Super-Kamiokande:2023jbt}%
  \BibitemOpen
  \bibfield  {author} {\bibinfo {author} {\bibfnamefont {K.}~\bibnamefont {Abe}} \emph {et~al.} (\bibinfo {collaboration} {Super-Kamiokande}),\ }\href {https://doi.org/10.1103/PhysRevD.109.092001} {\bibfield  {journal} {\bibinfo  {journal} {Phys. Rev. D}\ }\textbf {\bibinfo {volume} {109}},\ \bibinfo {pages} {092001} (\bibinfo {year} {2024})},\ \Eprint {https://arxiv.org/abs/2312.12907} {arXiv:2312.12907 [hep-ex]} \BibitemShut {NoStop}%
\bibitem [{\citenamefont {Agostini}\ \emph {et~al.}(2020{\natexlab{b}})\citenamefont {Agostini} \emph {et~al.}}]{Borexino:2017uhp}%
  \BibitemOpen
  \bibfield  {author} {\bibinfo {author} {\bibfnamefont {M.}~\bibnamefont {Agostini}} \emph {et~al.} (\bibinfo {collaboration} {Borexino}),\ }\href {https://doi.org/10.1103/PhysRevD.101.062001} {\bibfield  {journal} {\bibinfo  {journal} {Phys. Rev. D}\ }\textbf {\bibinfo {volume} {101}},\ \bibinfo {pages} {062001} (\bibinfo {year} {2020}{\natexlab{b}})},\ \Eprint {https://arxiv.org/abs/1709.00756} {arXiv:1709.00756 [hep-ex]} \BibitemShut {NoStop}%
\bibitem [{\citenamefont {Zhang}\ \emph {et~al.}(2016)\citenamefont {Zhang} \emph {et~al.}}]{Super-Kamiokande:2015xra}%
  \BibitemOpen
  \bibfield  {author} {\bibinfo {author} {\bibfnamefont {Y.}~\bibnamefont {Zhang}} \emph {et~al.} (\bibinfo {collaboration} {Super-Kamiokande}),\ }\href {https://doi.org/10.1103/PhysRevD.93.012004} {\bibfield  {journal} {\bibinfo  {journal} {Phys. Rev. D}\ }\textbf {\bibinfo {volume} {93}},\ \bibinfo {pages} {012004} (\bibinfo {year} {2016})},\ \Eprint {https://arxiv.org/abs/1509.08168} {arXiv:1509.08168 [hep-ex]} \BibitemShut {NoStop}%
\bibitem [{\citenamefont {Aharmim}\ \emph {et~al.}(2019)\citenamefont {Aharmim} \emph {et~al.}}]{SNO:2019pzy}%
  \BibitemOpen
  \bibfield  {author} {\bibinfo {author} {\bibfnamefont {B.}~\bibnamefont {Aharmim}} \emph {et~al.} (\bibinfo {collaboration} {SNO}),\ }\href {https://doi.org/10.1103/PhysRevD.100.112005} {\bibfield  {journal} {\bibinfo  {journal} {Phys. Rev. D}\ }\textbf {\bibinfo {volume} {100}},\ \bibinfo {pages} {112005} (\bibinfo {year} {2019})},\ \Eprint {https://arxiv.org/abs/1909.11728} {arXiv:1909.11728 [hep-ex]} \BibitemShut {NoStop}%
\bibitem [{\citenamefont {Mei}\ and\ \citenamefont {Hime}(2006)}]{Mei:2005gm}%
  \BibitemOpen
  \bibfield  {author} {\bibinfo {author} {\bibfnamefont {D.}~\bibnamefont {Mei}}\ and\ \bibinfo {author} {\bibfnamefont {A.}~\bibnamefont {Hime}},\ }\href {https://doi.org/10.1103/PhysRevD.73.053004} {\bibfield  {journal} {\bibinfo  {journal} {Phys. Rev. D}\ }\textbf {\bibinfo {volume} {73}},\ \bibinfo {pages} {053004} (\bibinfo {year} {2006})},\ \Eprint {https://arxiv.org/abs/astro-ph/0512125} {arXiv:astro-ph/0512125} \BibitemShut {NoStop}%
\bibitem [{\citenamefont {Wu}\ \emph {et~al.}(2023)\citenamefont {Wu} \emph {et~al.}}]{Wu:2022oxo}%
  \BibitemOpen
  \bibfield  {author} {\bibinfo {author} {\bibfnamefont {Y.}~\bibnamefont {Wu}} \emph {et~al.},\ }\href {https://doi.org/10.1016/j.nima.2023.168400} {\bibfield  {journal} {\bibinfo  {journal} {Nucl. Instrum. Meth. A}\ }\textbf {\bibinfo {volume} {1054}},\ \bibinfo {pages} {168400} (\bibinfo {year} {2023})},\ \Eprint {https://arxiv.org/abs/2212.13158} {arXiv:2212.13158 [hep-ex]} \BibitemShut {NoStop}%
\bibitem [{\citenamefont {Guo}\ \emph {et~al.}(2021)\citenamefont {Guo} \emph {et~al.}}]{JNE:2020bwn}%
  \BibitemOpen
  \bibfield  {author} {\bibinfo {author} {\bibfnamefont {Z.}~\bibnamefont {Guo}} \emph {et~al.} (\bibinfo {collaboration} {JNE}),\ }\href {https://doi.org/10.1088/1674-1137/abccae} {\bibfield  {journal} {\bibinfo  {journal} {Chin. Phys. C}\ }\textbf {\bibinfo {volume} {45}},\ \bibinfo {pages} {025001} (\bibinfo {year} {2021})},\ \Eprint {https://arxiv.org/abs/2007.15925} {arXiv:2007.15925 [physics.ins-det]} \BibitemShut {NoStop}%
\bibitem [{\citenamefont {Zhao}\ \emph {et~al.}(2022)\citenamefont {Zhao} \emph {et~al.}}]{JNE:2021cyb}%
  \BibitemOpen
  \bibfield  {author} {\bibinfo {author} {\bibfnamefont {L.}~\bibnamefont {Zhao}} \emph {et~al.} (\bibinfo {collaboration} {JNE}),\ }\href {https://doi.org/10.1088/1674-1137/ac66cc} {\bibfield  {journal} {\bibinfo  {journal} {Chin. Phys. C}\ }\textbf {\bibinfo {volume} {46}},\ \bibinfo {pages} {085001} (\bibinfo {year} {2022})},\ \Eprint {https://arxiv.org/abs/2108.04010} {arXiv:2108.04010 [hep-ex]} \BibitemShut {NoStop}%
\bibitem [{\citenamefont {Wang}\ \emph {et~al.}(2017)\citenamefont {Wang} \emph {et~al.}}]{Wang:2017ynm}%
  \BibitemOpen
  \bibfield  {author} {\bibinfo {author} {\bibfnamefont {Z.}~\bibnamefont {Wang}} \emph {et~al.},\ }\href {https://doi.org/10.1016/j.nima.2017.03.007} {\bibfield  {journal} {\bibinfo  {journal} {Nucl. Instrum. Meth. A}\ }\textbf {\bibinfo {volume} {855}},\ \bibinfo {pages} {81} (\bibinfo {year} {2017})},\ \Eprint {https://arxiv.org/abs/1703.01478} {arXiv:1703.01478 [physics.ins-det]} \BibitemShut {NoStop}%
\bibitem [{\citenamefont {Guo}\ \emph {et~al.}(2019)\citenamefont {Guo} \emph {et~al.}}]{Guo:2017nnr}%
  \BibitemOpen
  \bibfield  {author} {\bibinfo {author} {\bibfnamefont {Z.}~\bibnamefont {Guo}} \emph {et~al.},\ }\href {https://doi.org/10.1016/j.astropartphys.2019.02.001} {\bibfield  {journal} {\bibinfo  {journal} {Astropart. Phys.}\ }\textbf {\bibinfo {volume} {109}},\ \bibinfo {pages} {33} (\bibinfo {year} {2019})},\ \Eprint {https://arxiv.org/abs/1708.07781} {arXiv:1708.07781 [physics.ins-det]} \BibitemShut {NoStop}%
\bibitem [{\citenamefont {Luo}\ \emph {et~al.}(2023)\citenamefont {Luo} \emph {et~al.}}]{Luo:2022xrd}%
  \BibitemOpen
  \bibfield  {author} {\bibinfo {author} {\bibfnamefont {W.}~\bibnamefont {Luo}} \emph {et~al.},\ }\href {https://doi.org/10.1088/1748-0221/18/02/P02004} {\bibfield  {journal} {\bibinfo  {journal} {JINST}\ }\textbf {\bibinfo {volume} {18}}\bibfield  {number} {\bibinfo  {number} { (02)},\ \bibinfo {pages} {P02004}},\ }\Eprint {https://arxiv.org/abs/2209.13772} {arXiv:2209.13772 [physics.ins-det]} \BibitemShut {NoStop}%
\bibitem [{\citenamefont {Agostini}\ \emph {et~al.}(2022{\natexlab{a}})\citenamefont {Agostini} \emph {et~al.}}]{BOREXINO:2021xzc}%
  \BibitemOpen
  \bibfield  {author} {\bibinfo {author} {\bibfnamefont {M.}~\bibnamefont {Agostini}} \emph {et~al.} (\bibinfo {collaboration} {Borexino}),\ }\href {https://doi.org/10.1103/PhysRevD.105.052002} {\bibfield  {journal} {\bibinfo  {journal} {Phys. Rev. D}\ }\textbf {\bibinfo {volume} {105}},\ \bibinfo {pages} {052002} (\bibinfo {year} {2022}{\natexlab{a}})},\ \Eprint {https://arxiv.org/abs/2109.04770} {arXiv:2109.04770 [hep-ex]} \BibitemShut {NoStop}%
\bibitem [{\citenamefont {Agostini}\ \emph {et~al.}(2022{\natexlab{b}})\citenamefont {Agostini} \emph {et~al.}}]{BOREXINO:2021efb}%
  \BibitemOpen
  \bibfield  {author} {\bibinfo {author} {\bibfnamefont {M.}~\bibnamefont {Agostini}} \emph {et~al.} (\bibinfo {collaboration} {Borexino}),\ }\href {https://doi.org/10.1103/PhysRevLett.128.091803} {\bibfield  {journal} {\bibinfo  {journal} {Phys. Rev. Lett.}\ }\textbf {\bibinfo {volume} {128}},\ \bibinfo {pages} {091803} (\bibinfo {year} {2022}{\natexlab{b}})},\ \Eprint {https://arxiv.org/abs/2112.11816} {arXiv:2112.11816 [hep-ex]} \BibitemShut {NoStop}%
\bibitem [{\citenamefont {Allega}\ \emph {et~al.}(2024)\citenamefont {Allega} \emph {et~al.}}]{SNO:2023cnz}%
  \BibitemOpen
  \bibfield  {author} {\bibinfo {author} {\bibfnamefont {A.}~\bibnamefont {Allega}} \emph {et~al.} (\bibinfo {collaboration} {SNO+}),\ }\href {https://doi.org/10.1103/PhysRevD.109.072002} {\bibfield  {journal} {\bibinfo  {journal} {Phys. Rev. D}\ }\textbf {\bibinfo {volume} {109}},\ \bibinfo {pages} {072002} (\bibinfo {year} {2024})},\ \Eprint {https://arxiv.org/abs/2309.06341} {arXiv:2309.06341 [hep-ex]} \BibitemShut {NoStop}%
\bibitem [{\citenamefont {Agostinelli}\ \emph {et~al.}(2003)\citenamefont {Agostinelli} \emph {et~al.}}]{GEANT4:2002zbu}%
  \BibitemOpen
  \bibfield  {author} {\bibinfo {author} {\bibfnamefont {S.}~\bibnamefont {Agostinelli}} \emph {et~al.} (\bibinfo {collaboration} {GEANT4}),\ }\href {https://doi.org/10.1016/S0168-9002(03)01368-8} {\bibfield  {journal} {\bibinfo  {journal} {Nucl. Instrum. Meth. A}\ }\textbf {\bibinfo {volume} {506}},\ \bibinfo {pages} {250} (\bibinfo {year} {2003})}\BibitemShut {NoStop}%
\bibitem [{\citenamefont {Allison}\ \emph {et~al.}(2006)\citenamefont {Allison} \emph {et~al.}}]{allison2006geant4}%
  \BibitemOpen
  \bibfield  {author} {\bibinfo {author} {\bibfnamefont {J.}~\bibnamefont {Allison}} \emph {et~al.},\ }\href {https://doi.org/10.1109/TNS.2006.869826} {\bibfield  {journal} {\bibinfo  {journal} {IEEE Trans. Nucl. Sci.}\ }\textbf {\bibinfo {volume} {53}},\ \bibinfo {pages} {270} (\bibinfo {year} {2006})}\BibitemShut {NoStop}%
\bibitem [{\citenamefont {Guan}\ \emph {et~al.}(2015)\citenamefont {Guan}, \citenamefont {Chu}, \citenamefont {Cao}, \citenamefont {Luk},\ and\ \citenamefont {Yang}}]{guan2015parametrizationcosmicraymuonflux}%
  \BibitemOpen
  \bibfield  {author} {\bibinfo {author} {\bibfnamefont {M.}~\bibnamefont {Guan}}, \bibinfo {author} {\bibfnamefont {M.-C.}\ \bibnamefont {Chu}}, \bibinfo {author} {\bibfnamefont {J.}~\bibnamefont {Cao}}, \bibinfo {author} {\bibfnamefont {K.-B.}\ \bibnamefont {Luk}},\ and\ \bibinfo {author} {\bibfnamefont {C.}~\bibnamefont {Yang}},\ }\href {https://arxiv.org/abs/1509.06176} {\bibinfo {title} {A parametrization of the cosmic-ray muon flux at sea-level}} (\bibinfo {year} {2015}),\ \Eprint {https://arxiv.org/abs/1509.06176} {arXiv:1509.06176 [hep-ex]} \BibitemShut {NoStop}%
\bibitem [{\citenamefont {Fedynitch}\ \emph {et~al.}(2015)\citenamefont {Fedynitch}, \citenamefont {Engel}, \citenamefont {Gaisser}, \citenamefont {Riehn},\ and\ \citenamefont {Stanev}}]{Fedynitch:2015zma}%
  \BibitemOpen
  \bibfield  {author} {\bibinfo {author} {\bibfnamefont {A.}~\bibnamefont {Fedynitch}}, \bibinfo {author} {\bibfnamefont {R.}~\bibnamefont {Engel}}, \bibinfo {author} {\bibfnamefont {T.~K.}\ \bibnamefont {Gaisser}}, \bibinfo {author} {\bibfnamefont {F.}~\bibnamefont {Riehn}},\ and\ \bibinfo {author} {\bibfnamefont {T.}~\bibnamefont {Stanev}},\ }\href {https://doi.org/10.1051/epjconf/20159908001} {\bibfield  {journal} {\bibinfo  {journal} {EPJ Web Conf.}\ }\textbf {\bibinfo {volume} {99}},\ \bibinfo {pages} {08001} (\bibinfo {year} {2015})},\ \Eprint {https://arxiv.org/abs/1503.00544} {arXiv:1503.00544 [hep-ph]} \BibitemShut {NoStop}%
\bibitem [{\citenamefont {Fedynitch}\ \emph {et~al.}(2019)\citenamefont {Fedynitch}, \citenamefont {Riehn}, \citenamefont {Engel}, \citenamefont {Gaisser},\ and\ \citenamefont {Stanev}}]{Fedynitch:2018cbl}%
  \BibitemOpen
  \bibfield  {author} {\bibinfo {author} {\bibfnamefont {A.}~\bibnamefont {Fedynitch}}, \bibinfo {author} {\bibfnamefont {F.}~\bibnamefont {Riehn}}, \bibinfo {author} {\bibfnamefont {R.}~\bibnamefont {Engel}}, \bibinfo {author} {\bibfnamefont {T.~K.}\ \bibnamefont {Gaisser}},\ and\ \bibinfo {author} {\bibfnamefont {T.}~\bibnamefont {Stanev}},\ }\href {https://doi.org/10.1103/PhysRevD.100.103018} {\bibfield  {journal} {\bibinfo  {journal} {Phys. Rev. D}\ }\textbf {\bibinfo {volume} {100}},\ \bibinfo {pages} {103018} (\bibinfo {year} {2019})},\ \Eprint {https://arxiv.org/abs/1806.04140} {arXiv:1806.04140 [hep-ph]} \BibitemShut {NoStop}%
\bibitem [{\citenamefont {Farr}\ \emph {et~al.}(2007)\citenamefont {Farr} \emph {et~al.}}]{terrain2007}%
  \BibitemOpen
  \bibfield  {author} {\bibinfo {author} {\bibfnamefont {T.~G.}\ \bibnamefont {Farr}} \emph {et~al.},\ }\href {https://doi.org/10.1029/2005RG000183} {\bibfield  {journal} {\bibinfo  {journal} {Reviews of Geophysics}\ }\textbf {\bibinfo {volume} {47}} (\bibinfo {year} {2007})}\BibitemShut {NoStop}%
\bibitem [{\citenamefont {Zheng}\ \emph {et~al.}(2024)\citenamefont {Zheng}, \citenamefont {Li}, \citenamefont {Feng}, \citenamefont {Xu},\ and\ \citenamefont {Xiao}}]{zheng2024three}%
  \BibitemOpen
  \bibfield  {author} {\bibinfo {author} {\bibfnamefont {M.}~\bibnamefont {Zheng}}, \bibinfo {author} {\bibfnamefont {S.}~\bibnamefont {Li}}, \bibinfo {author} {\bibfnamefont {Z.}~\bibnamefont {Feng}}, \bibinfo {author} {\bibfnamefont {H.}~\bibnamefont {Xu}},\ and\ \bibinfo {author} {\bibfnamefont {Y.}~\bibnamefont {Xiao}},\ }\href {https://doi.org/https://doi.org/10.1016/j.ijmst.2023.12.007} {\bibfield  {journal} {\bibinfo  {journal} {International Journal of Mining Science and Technology}\ }\textbf {\bibinfo {volume} {34}},\ \bibinfo {pages} {179} (\bibinfo {year} {2024})}\BibitemShut {NoStop}%
\bibitem [{\citenamefont {Haynes}(2016)}]{earth2016}%
  \BibitemOpen
  \bibfield  {author} {\bibinfo {author} {\bibfnamefont {W.~M.}\ \bibnamefont {Haynes}},\ }\href {https://doi.org/10.1201/9781315380476} {\emph {\bibinfo {title} {CRC Handbook of Chemistry and Physics}}},\ Vol.~\bibinfo {volume} {97}\ (\bibinfo  {publisher} {CRC Press},\ \bibinfo {year} {2016})\BibitemShut {NoStop}%
\bibitem [{\citenamefont {Li}\ and\ \citenamefont {Beacom}(2015)}]{Li:2015kpa}%
  \BibitemOpen
  \bibfield  {author} {\bibinfo {author} {\bibfnamefont {S.~W.}\ \bibnamefont {Li}}\ and\ \bibinfo {author} {\bibfnamefont {J.~F.}\ \bibnamefont {Beacom}},\ }\href {https://doi.org/10.1103/PhysRevD.91.105005} {\bibfield  {journal} {\bibinfo  {journal} {Phys. Rev. D}\ }\textbf {\bibinfo {volume} {91}},\ \bibinfo {pages} {105005} (\bibinfo {year} {2015})},\ \Eprint {https://arxiv.org/abs/1503.04823} {arXiv:1503.04823 [hep-ph]} \BibitemShut {NoStop}%
\bibitem [{\citenamefont {Bellamy}\ \emph {et~al.}(1994)\citenamefont {Bellamy} \emph {et~al.}}]{bellamy1994absolute}%
  \BibitemOpen
  \bibfield  {author} {\bibinfo {author} {\bibfnamefont {E.~H.}\ \bibnamefont {Bellamy}} \emph {et~al.},\ }\href {https://doi.org/10.1016/0168-9002(94)90183-X} {\bibfield  {journal} {\bibinfo  {journal} {Nucl. Instrum. Meth. A}\ }\textbf {\bibinfo {volume} {339}},\ \bibinfo {pages} {468} (\bibinfo {year} {1994})}\BibitemShut {NoStop}%
\bibitem [{\citenamefont {An}\ \emph {et~al.}(2017)\citenamefont {An} \emph {et~al.}}]{an2017measurement}%
  \BibitemOpen
  \bibfield  {author} {\bibinfo {author} {\bibfnamefont {F.~P.}\ \bibnamefont {An}} \emph {et~al.} (\bibinfo {collaboration} {Daya Bay}),\ }\href {https://doi.org/10.1103/PhysRevD.95.072006} {\bibfield  {journal} {\bibinfo  {journal} {Phys. Rev. D}\ }\textbf {\bibinfo {volume} {95}},\ \bibinfo {pages} {072006} (\bibinfo {year} {2017})},\ \Eprint {https://arxiv.org/abs/1610.04802} {arXiv:1610.04802 [hep-ex]} \BibitemShut {NoStop}%
\bibitem [{\citenamefont {Zhang}\ \emph {et~al.}(2022)\citenamefont {Zhang}, \citenamefont {Wang},\ and\ \citenamefont {Chen}}]{zhang2022mou}%
  \BibitemOpen
  \bibfield  {author} {\bibinfo {author} {\bibfnamefont {B.}~\bibnamefont {Zhang}}, \bibinfo {author} {\bibfnamefont {Z.}~\bibnamefont {Wang}},\ and\ \bibinfo {author} {\bibfnamefont {S.}~\bibnamefont {Chen}},\ }\bibfield  {journal} {\bibinfo  {journal} {Applied Sciences}\ }\textbf {\bibinfo {volume} {12}},\ \href {https://doi.org/10.3390/app122110975} {10.3390/app122110975} (\bibinfo {year} {2022}),\ \Eprint {https://arxiv.org/abs/2209.11974} {arXiv:2209.11974} \BibitemShut {NoStop}%
\bibitem [{\citenamefont {Cheng}\ \emph {et~al.}(2016)\citenamefont {Cheng} \emph {et~al.}}]{cheng2016determination}%
  \BibitemOpen
  \bibfield  {author} {\bibinfo {author} {\bibfnamefont {J.-H.}\ \bibnamefont {Cheng}} \emph {et~al.},\ }\href {https://doi.org/10.1016/j.nima.2016.05.010} {\bibfield  {journal} {\bibinfo  {journal} {Nucl. Instrum. Meth. A}\ }\textbf {\bibinfo {volume} {827}},\ \bibinfo {pages} {165} (\bibinfo {year} {2016})},\ \Eprint {https://arxiv.org/abs/1603.04433} {arXiv:1603.04433 [physics.ins-det]} \BibitemShut {NoStop}%
\bibitem [{\citenamefont {An}\ \emph {et~al.}(2018)\citenamefont {An} \emph {et~al.}}]{DayaBay:2017txw}%
  \BibitemOpen
  \bibfield  {author} {\bibinfo {author} {\bibfnamefont {F.~P.}\ \bibnamefont {An}} \emph {et~al.} (\bibinfo {collaboration} {Daya Bay}),\ }\href {https://doi.org/10.1103/PhysRevD.97.052009} {\bibfield  {journal} {\bibinfo  {journal} {Phys. Rev. D}\ }\textbf {\bibinfo {volume} {97}},\ \bibinfo {pages} {052009} (\bibinfo {year} {2018})},\ \Eprint {https://arxiv.org/abs/1711.00588} {arXiv:1711.00588 [hep-ex]} \BibitemShut {NoStop}%
\bibitem [{\citenamefont {Wang}\ \emph {et~al.}(2001)\citenamefont {Wang}, \citenamefont {Balic}, \citenamefont {Gratta}, \citenamefont {Fasso}, \citenamefont {Roesler},\ and\ \citenamefont {Ferrari}}]{Wang:2001fq}%
  \BibitemOpen
  \bibfield  {author} {\bibinfo {author} {\bibfnamefont {Y.~F.}\ \bibnamefont {Wang}}, \bibinfo {author} {\bibfnamefont {V.}~\bibnamefont {Balic}}, \bibinfo {author} {\bibfnamefont {G.}~\bibnamefont {Gratta}}, \bibinfo {author} {\bibfnamefont {A.}~\bibnamefont {Fasso}}, \bibinfo {author} {\bibfnamefont {S.}~\bibnamefont {Roesler}},\ and\ \bibinfo {author} {\bibfnamefont {A.}~\bibnamefont {Ferrari}},\ }\href {https://doi.org/10.1103/PhysRevD.64.013012} {\bibfield  {journal} {\bibinfo  {journal} {Phys. Rev. D}\ }\textbf {\bibinfo {volume} {64}},\ \bibinfo {pages} {013012} (\bibinfo {year} {2001})},\ \Eprint {https://arxiv.org/abs/hep-ex/0101049} {arXiv:hep-ex/0101049} \BibitemShut {NoStop}%
\bibitem [{\citenamefont {Hertenberger}\ \emph {et~al.}(1995)\citenamefont {Hertenberger}, \citenamefont {Chen},\ and\ \citenamefont {Dougherty}}]{Hertenberger:1995ae}%
  \BibitemOpen
  \bibfield  {author} {\bibinfo {author} {\bibfnamefont {R.}~\bibnamefont {Hertenberger}}, \bibinfo {author} {\bibfnamefont {M.}~\bibnamefont {Chen}},\ and\ \bibinfo {author} {\bibfnamefont {B.~L.}\ \bibnamefont {Dougherty}},\ }\href {https://doi.org/10.1103/PhysRevC.52.3449} {\bibfield  {journal} {\bibinfo  {journal} {Phys. Rev. C}\ }\textbf {\bibinfo {volume} {52}},\ \bibinfo {pages} {3449} (\bibinfo {year} {1995})}\BibitemShut {NoStop}%
\bibitem [{\citenamefont {Boehm}\ \emph {et~al.}(2000)\citenamefont {Boehm} \emph {et~al.}}]{Boehm:2000ru}%
  \BibitemOpen
  \bibfield  {author} {\bibinfo {author} {\bibfnamefont {F.}~\bibnamefont {Boehm}} \emph {et~al.},\ }\href {https://doi.org/10.1103/PhysRevD.62.092005} {\bibfield  {journal} {\bibinfo  {journal} {Phys. Rev. D}\ }\textbf {\bibinfo {volume} {62}},\ \bibinfo {pages} {092005} (\bibinfo {year} {2000})},\ \Eprint {https://arxiv.org/abs/hep-ex/0006014} {arXiv:hep-ex/0006014} \BibitemShut {NoStop}%
\bibitem [{\citenamefont {Blyth}\ \emph {et~al.}(2016)\citenamefont {Blyth} \emph {et~al.}}]{AberdeenTunnelExperiment:2015uaa}%
  \BibitemOpen
  \bibfield  {author} {\bibinfo {author} {\bibfnamefont {S.~C.}\ \bibnamefont {Blyth}} \emph {et~al.} (\bibinfo {collaboration} {Aberdeen Tunnel Experiment}),\ }\href {https://doi.org/10.1103/PhysRevD.93.072005} {\bibfield  {journal} {\bibinfo  {journal} {Phys. Rev. D}\ }\textbf {\bibinfo {volume} {93}},\ \bibinfo {pages} {072005} (\bibinfo {year} {2016})},\ \Eprint {https://arxiv.org/abs/1509.09038} {arXiv:1509.09038 [physics.ins-det]} \BibitemShut {NoStop}%
\bibitem [{\citenamefont {Abe}\ \emph {et~al.}(2010)\citenamefont {Abe} \emph {et~al.}}]{KamLAND:2009zwo}%
  \BibitemOpen
  \bibfield  {author} {\bibinfo {author} {\bibfnamefont {S.}~\bibnamefont {Abe}} \emph {et~al.} (\bibinfo {collaboration} {KamLAND}),\ }\href {https://doi.org/10.1103/PhysRevC.81.025807} {\bibfield  {journal} {\bibinfo  {journal} {Phys. Rev. C}\ }\textbf {\bibinfo {volume} {81}},\ \bibinfo {pages} {025807} (\bibinfo {year} {2010})},\ \Eprint {https://arxiv.org/abs/0907.0066} {arXiv:0907.0066 [hep-ex]} \BibitemShut {NoStop}%
\bibitem [{\citenamefont {Aglietta}\ \emph {et~al.}(1999)\citenamefont {Aglietta} \emph {et~al.}}]{LVD:1999ezh}%
  \BibitemOpen
  \bibfield  {author} {\bibinfo {author} {\bibfnamefont {M.}~\bibnamefont {Aglietta}} \emph {et~al.} (\bibinfo {collaboration} {LVD}),\ }in\ \href@noop {} {\emph {\bibinfo {booktitle} {{26th International Cosmic Ray Conference}}}}\ (\bibinfo {year} {1999})\ \Eprint {https://arxiv.org/abs/hep-ex/9905047} {arXiv:hep-ex/9905047} \BibitemShut {NoStop}%
\bibitem [{\citenamefont {Bellini}\ \emph {et~al.}(2013)\citenamefont {Bellini} \emph {et~al.}}]{Borexino:2013cke}%
  \BibitemOpen
  \bibfield  {author} {\bibinfo {author} {\bibfnamefont {G.}~\bibnamefont {Bellini}} \emph {et~al.} (\bibinfo {collaboration} {Borexino}),\ }\href {https://doi.org/10.1088/1475-7516/2013/08/049} {\bibfield  {journal} {\bibinfo  {journal} {JCAP}\ }\textbf {\bibinfo {volume} {08}},\ \bibinfo {pages} {049}},\ \Eprint {https://arxiv.org/abs/1304.7381} {arXiv:1304.7381 [physics.ins-det]} \BibitemShut {NoStop}%
\end{thebibliography}%

\end{document}